\documentclass[fleqn,usenatbib]{mnras}

\usepackage[T1]{fontenc}
\usepackage{ae,aecompl}

\usepackage{graphicx}	
\usepackage{amsmath}	
\usepackage{amssymb}	
\usepackage[dvipsnames]{xcolor}

\usepackage{amsmath}
\newcommand{\HA}{{\ensuremath{\rm H\alpha}}}
\newcommand{\HB}{{\ensuremath{\rm H\beta}}}
\newcommand{\HG}{{\ensuremath{\rm H\gamma}}}

\newcommand{\HII}{{\ensuremath{\rm H\text{\small II}}}}
\newcommand{\HI}{\ensuremath{\rm H\text{\small I}}}

\newcommand\pfrac[2]{\ensuremath{\left(\frac{#1}{#2}\right)}}
\newcommand{\surf}[1]{\ensuremath{\rm #1 mag/{_{\Box^{''}}}}}
\newcommand{\llha}[1]{{\ensuremath{{\rm log}L_\HA{}#1}}}
\newcommand{\lcomp}{{\ensuremath{L_{c}}}}

 \newcommand\nobrkhyph{\mbox{-}}
\newcommand{\band}[1]{#1\nobrkhyph{}band}

\newcommand{\arcsecond}[2]{\ensuremath{#1''\!\!.#2}}

\makeatletter
\newcommand{\Rom}[1]{\expandafter\@slowromancap\romannumeral #1@}
\makeatother
\newcommand{\ionl}[3]{[#1\ensuremath{\,}{\small \Rom{#2}}]\relax\ensuremath{\lambda}#3}

\newcommand{\figref}[1]{Figure~\ref{fig:#1}}
\newcommand{\secref}[1]{Section~\ref{sec:#1}}
\newcommand{\tblref}[1]{Table~\ref{tbl:#1}}
\newcommand{\equref}[1]{Equation~\ref{equ:#1}}

\def\apj{ApJ}
\def\aap{A\&A}
\def\apjl{ApJ}
\def\aj{AJ}

\def\mnras{MNRAS}

\def\apjs{ApJS}

\title[The Hyper-Stable Disc of UGC 8839]{The Hyper-Stable Disc of UGC 8839}

\author[J.E. Young and M. Eleazer]{
Jason E. Young$^{1}$\thanks{E-mail: jyoung@mtholyoke.edu}
and Miriam Eleazer$^{1,2}$\thanks{E-mail: meleazer@wesleyan.edu}
\\
$^{1}$Astronomy Department, Mount Holyoke College, 50 College Street, 01075, United States of America\\
$^{2}$Astronomy Department, Wesleyan University, 45 Wyllys Avenue, 06459, United States of America\\
}

\date{Accepted XXX. Received YYY; in original form ZZZ}

\pubyear{2015}

\begin{document}
\label{firstpage}
\pagerange{\pageref{firstpage}--\pageref{lastpage}}
\maketitle

\begin{abstract}
The low surface brightness (LSB) spiral UGC~8839 is nearly devoid of star formation aside from a large \HII{} region complex located in the extreme outer disc. In order to understand the origin and nature of this complex, we compare new \HA{} and archival \band{broad} images of UGC~8839 to similar data for four other spiral galaxies. We conclude that the extreme off-axis star formation in UGC~8839 is likely due to a hyper-stable disk that is dark matter dominated at all radii, with the Toomre parameter reaching a minimum only in the extreme outer disc. Using analysis strategies designed to be particularly insensitive to the pitfalls of low-surface brightness objects and small number statistics, we determine that the presence of this complex in UGC~8839 is not exceptional when the \HII{} region luminosity function is modelled by a power law, suggesting that it is a native structure and not a merging satellite. However, we find that the entire population of \HII{} regions in UGC~8839 shows a preference for larger galactocentric radii when compared to \HII{} regions in the other galaxies in our sample.  UGC~8839 dramatically highlights the relationship between the baryonic/dark matter ratio and disk stability. A three-body interaction, similar to a scaled-down version of the interaction suspected to be responsible for Malin~1, is consistent with the extreme outer disk star formation that we see in the extended disk of UGC~8839.

\end{abstract}



\section{Introduction}
\label{sec:intro}

Low surface brightness (LSB) spirals are extreme late-type spirals (Sd/Sm/Irr) with blue optical colours and high gas fractions \citet{McGaugh1995morphology,deBlok1995,deBlok1996}. Paradoxically, they also typically exhibit low star-formation rates \citep[e.g.,][]{vandenHoek2000,Kim2007}. This presents an interesting problem for our understanding of the ISM as the bidirectional interface between galaxy dynamics and stellar populations. As a particularly gas-rich but \HA{}-faint object, the LSB spiral UGC~8839 is an excellent test case for ideas about the evolution of LSB spirals.

This paradox may be partly solved by sporadic star formation. \citet{Boissier2008} find that LSB spirals have Galex FUV-NUV colours similar to but slightly redder than HSB spirals, and discuss the possibility that the red FUV-NUV colours are due to the fading of young but not actively forming population. Using hydrodynamic simulations \citet{Vorobyov2009} are able to replicate typical B-V colours and \HA{} equivalent widths of LSB galaxies with sporadic star formation; irregular bursts keep the galaxy disc comparatively blue, but the time-averaged star formation remains low. In our earlier paper, \citet{Young2020}, we reported on the spatially resolved star-formation history of the LSB spiral UGC~628, and found evidence that recent star formation seems to be sporadic, occasionally rising to a starburst level.

However, we also found that the current/recent star formation in UGC~628 is located almost entirely on the outer edge of the disc. This may be typical of LSB spirals: In a sample of 1000 galaxies, \citet{Huang2013} showed that edge-dominant starbursts are more common in LSB or otherwise late-type galaxies, although few are as edge-dominated as UGC~628.

Edge-dominant star formation may also explain the comparatively flat metallicity gradients reported in LSB galaxies. \citet{deBlok1998I} measure O/H ratios using nebular lines in a sample of three LSB galaxies, and find very shallow and/or flat metallicity gradients.  In \citet{Young2015} we report a slightly inverted metallicity gradient in UGC~628. \citet{Bresolin2015} find a plausible solution by pointing out that LSB spirals also have shallower stellar light profiles; that is, longer exponential scale lengths for their exponential discs. When metallicity gradients are calculated in terms of scale length instead of physical units, HSB and LSB spirals have similar metallicity gradients. If LSB spirals preferentially form stars near edges of their discs, this would result in discs where both the stars and the metals are more diffuse, and cleanly explain the longer scale lengths and shallower metallicity profiles.

If indeed the characteristics of LSB spirals can be largely explained by sporadic, edge-dominant star formation, then isolated LSB spirals that show extreme off-centre star formation are cases that highlight the physical differences between LSB and HSB spirals.

We can also gain some insight into the physical conditions which inhibit star formation in gas-rich LSB spirals by looking at the \HII{} region luminosity function (HRLF). Classical \HII{} / morphology relationships \cite[e.g.,][]{KEH1989} show that the HRLF follows a steep power law in early-type spirals and a shallow power law in late-type spirals. That is, early types favour many small \HII{} regions while late types favour a small number of larger \HII{} regions. As extreme late-types, LSB spirals might be expected to have a shallow HRLF. However, unlike most high surface brightness (HSB) late types, LSB spirals typically have low star-formation rates, hampering attempts to measure the HRLF in individual LSB spirals.

\citet{Helmboldt2009} tackle this problem by stacking HRLFs from LSB and HSB spirals, and then comparing the stacked HRLFs. They find that a key difference  is that the HRLF in HSB spirals is a Schechter function, which becomes steeper above \llha{\rm [erg\,s^{-1}]\!=\! 38.6}. LSB spirals do not seem to show this break. Overall low star-formation rates mean that LSB spirals are less likely to host giant \HII{} regions, but, per unit star-formation rate, they are actually more likely to host giant \HII{} regions. If this is correct, then giant \HII{} regions with \llha{{\rm [erg\,s^{-1}]}{\,\gtrsim 38.6}} may be the sites of greatest contrast between HSB and LSB spirals, and making cases such as UGC~8839 particularly worth studying.

Given these facts, the LSB spiral UGC~8839 is a very interesting case. With a surface brightness of $\mu_g(0)=23.1\surf{}$, it falls well within the LSB category, though it is by no means an extreme example. We present \HA{} images which show that UGC~8839 is largely devoid of star formation, with the exception of a \llha{{\rm [erg\,s^{-1}]}{=\!39}} \HII{} region complex located at the extreme outer edge of the galactic disc.  In this work, we examine the regions of this complex in the context of the HRLF and the broad-band disc of UGC~8839. For comparison, we examine three other LSB spirals and an HSB spiral in the same fashion.

In \secref{observations} we describe our targets list, our observations, and our reduction methods. In \secref{luminosity} we develop a methodology to determine a probability distribution for the power-law index of the HRLF in the limit of a small sample size. We then use the constraints on the power-law index to show that the giant \HII{} region complex in UGC~8839 is likely drawn from the same population as the remaining smaller \HII{} regions in UGC~8839. In \secref{location} we show that the entire population of \HII{} regions in UGC~8839 is located further out in the galactic disc than the \HII{} regions in our other sample galaxies. In \secref{discussion} we examine several interpretations and implications for other LSB spirals.

\section{Observations}
\label{sec:observations}

\subsection{Targets}
\label{sec:targets}
Our targets are listed in \tblref{targets}, and include four LSB spirals and one HSB spiral. These targets were drawn from the catalogues of LSB galaxies presented in \cite{McGaugh1995morphology} and \cite{Kim2007}. Because earlier studies of the HRLF have achieved reasonably complete samples down to $\log{L_\HA{}{\rm[erg\,s^{-1}]}}=37$ in galaxies out to 30Mpc \citep[e.g.,][]{Caldwell1991}, we restricted our targets to distances $\lesssim 30\rm Mpc$. 

{\bf UGC 8839:} This galaxy is classified as Im, but close inspection reveals faint extended spiral arms (\figref{contours}). The central surface brightness of UGC~8839 is roughly a magnitude fainter than typical dark skies, making it decidedly an LSB spiral. The large \HII{} region complex comprised of regions A, B, and C (\figref{images}) appears to be an extension of the southern spiral arm, and accounts for approximately 69\% of the discrete star formation in this galaxy. We will refer to the complex collectively as ABC, although the individual regions will be analysed as discrete regions in the same manner as the other regions in our sample. The identification of this complex, located in the extreme outer disc, motivated this paper.

{\bf UGC 5633:} This is a strongly barred galaxy, a rare characteristic for LSB spirals, yet with a central surface brightness of $\mu_0(g)=\surf{22.9}$ UGC~5633 clearly falls within the LSB category. In \figref{images} we see some \HA{} emission associated with this bar, suggesting that it is a gaseous as well as a stellar bar.

{\bf UGC 6151:} The \HA{} image for this object shows spirals arms traced out by the \HII{} regions. One of the \HII{} regions is located at the exact centre of the disc. This central region is likely the cause of the SDSS DR15 spectral classification of UGC~6151 as a starburst galaxy; in fact the galaxy-wide \HA{} luminosity is not particularly high (see \tblref{targets}).

{\bf UGC 6181:} This galaxy is a marginal LSB, with $\mu_0(g)=\surf{22.2}$. It hosts numerous star-forming regions at the edge of the disc. The projected distance between UGC~6181 and UGC~6151 is $\sim\!640$kpc, close enough that they may be bound, although there are no obvious signs of interaction.

{\bf NGC 4455:} This HSB spiral is included in this study so that our sample would span a range of surface brightnesses and chosen specifically because of its proximity (only 10 Mpc) and availability during our observing run. It is a late type spiral with an SBd morphology, although the classification is dubious since it is inclined at $78^\circ$. The estimated mass is log($\mathcal{M_\star}/\mathcal{M_\odot}$) = 11 \citep{Kim2007}.



\begin{table*}
\caption{Target Galaxies and Observations}
\begin{tabular}{lcccccccccccccc}
	&	Type$^a$&	D$^a$	&	$\rm g$	&	$\rm M_g$	&	{$\mu_0(g)$}	&	$\rm r$	&	$\rm M_r$	&	{$\mu_0(r)$}	&	$\rm z$	&	$\rm M_z$	&	{$\mu_0(z)$}	&	\HA{}+0	&	\HA{}+16	&	Date\\
	&		&	Mpc	&	mag.	&	mag.	&	\surf{}	&	mag.	&	mag.	&	\surf{}	&	mag.	&	mag.	&	\surf{}	&	sec.	&	sec. & 03/2018\\		
\hline																												
UGC~8839	&	Im	&	22.2	&	14.6	&	-17.1	&	23.0	&	14.2	&	-17.5	&	22.6	&	14.0	&	-17.7	&	22.3	&	2120	&	1820	&	16,17\\
NGC~4455	&	SBd	&	10.4	&	12.8	&	-17.3	&	20.8	&	12.5	&	-17.6	&	20.4	&	12.3	&	-17.8	&	20.2	&	480	&	160	&	16\\
UGC~6181	&	Im	&	23.7	&	14.2	&	-17.7	&	22.2	&	13.8	&	-18.1	&	21.8	&	13.1	&	-18.8	&	21.5	&	3900	&	2400	&	16\\
UGC~6151	&	Sm	&	24.3	&	14.3	&	-17.6	&	22.8	&	14.0	&	-17.9	&	22.4	&	13.8	&	-18.1	&	22.1	&	3900	&	2400	&	16\\
UGC~5633	&	SBdm	&	23.4	&	14.1	&	-17.8	&	22.9	&	13.7	&	-18.2	&	22.4	&	13.5	&	-18.3	&	22.1	&	1620	&	1080	&	17\\

\hline
\multicolumn{5}{l}{$^a$ NED}\\
\label{tbl:targets}
\end{tabular}
\end{table*}

\subsection{Observations and Reduction}
\label{sec:reduction}

The \HA{} images in our study were collected with the Half Degree Imager (HDI) on the WIYN 0.9 meter telescope at Kitt Peak National Observatory on March 16-18 of 2018.  

\band{On} observations were made using the \HA+0nm filter ($\lambda=6547-6606$\AA{}), and \band{off} observations were made using the \HA+16nm filter ($\lambda=6712-6772$\AA{}). The data was reduced in a standard fashion using biases and a combination of dome and twilight flats. Because of the large field of view of the HDI, it was possible to identify SDSS quasars with publicly available spectra in all of our target fields, allowing for contemporaneous flux calibration. To do this, we convolved the SDSS quasar spectra with our filter transmission curves \citep{SDSSDR16}, and then adopted the average flux/count-rate ratios as our calibration and the scatter in the ratios as the uncertainties on our calibration.

As a final calibration step, the flux calibrated images were then corrected for foreground galactic extinction using the Cardelli extinction law \citep{Cardelli1989} and the E(B-V) values listed on NASA/IPEC Extragalactic Database (NED), derived from the dust maps in \cite{Schlegel1998}.

After calibration, the \band{off} \HA+16nm images were subtracted from the \band{on} \HA+0nm images to remove the stellar continuum. Then, 2\AA{} equivalent width of stellar continuum was added back into the \band{on} images to account for the stellar absorption underneath the \HA{} emission line. The choice of 2\AA{} equivalent width as an estimation of the strength of the Balmer absorption lines (in the absence of high resolution spectra) is motivated by the careful analysis of \HA{} images of \HII{} regions in \cite{McCall1985} and \cite{Oey1993}, of \HB{} images of LSB galaxies in \cite{KuziodeNaray2004}, and of \HB{} and \HG{} data in our earlier work, \cite{Young2015}. The average fractional correction to the \HA{} fluxes due to 2\AA{} of continuum absorption was $0.054^{+0.051}_{-0.027}$. The average fractional correction for regions in UGC~8839 was even smaller, $0.007^{+0.004}_{-0.003}$. Note that these corrections are generally much smaller than the measurement uncertainties listed in \tblref{regions}, and have relatively little impact on the analysis below. As discussed in \secref{halpha}, uncertainties in the continuum fluxes were propagated through the continuum removal and the 2\AA{} correction in quadrature with the uncertainties in the on-band \HA{} fluxes. No attempt was made to remove the \ionl{N}{2}{6584} line from the \HA{} images.

\begin{figure}
  \includegraphics[width=0.47\textwidth]{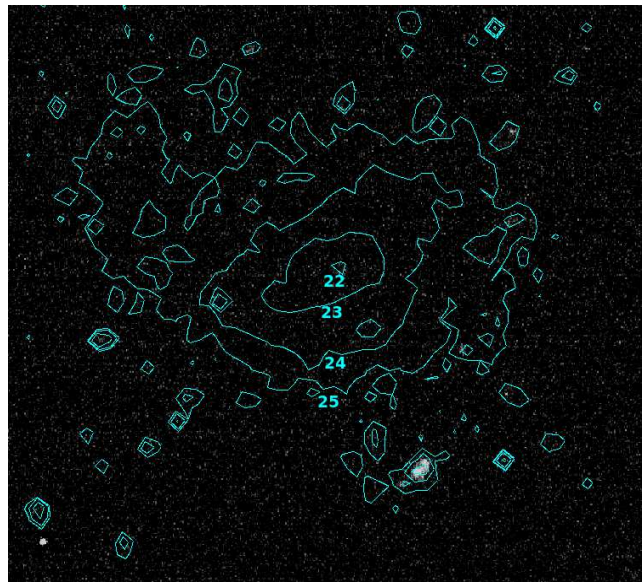}
  \caption{ \HA{} image of UGC~8839, overlaid with SDSS \band{g} surface brightness contours. The contours are labelled in units of \surf{}.}
  \label{fig:contours}
\end{figure}

In \figref{contours} we show the \HA{} image of UGC~8839, overlaid with SDSS DR16 \band{g} surface brightness contours \citep{SDSSDR16}. The \HA{} image is nearly blank, aside from the bright spot located in the lower right, the ABC complex. The identification of individual \HII{} regions will be discussed in \secref{halpha}. Curiously, UGC~8839 seems to be nearly devoid of star formation, aside from this large clump at the extreme edge of the disc. Although UGC~8839 is classified as an Im type, the \band{g} contours do show faint spiral arms. Interestingly, the ABC complex appears to be an extension of the south western arm.

\subsection{Identification and Measurement of \HII{} Regions}
\label{sec:halpha}

\begin{figure*}
  \includegraphics[width=0.85\textwidth]{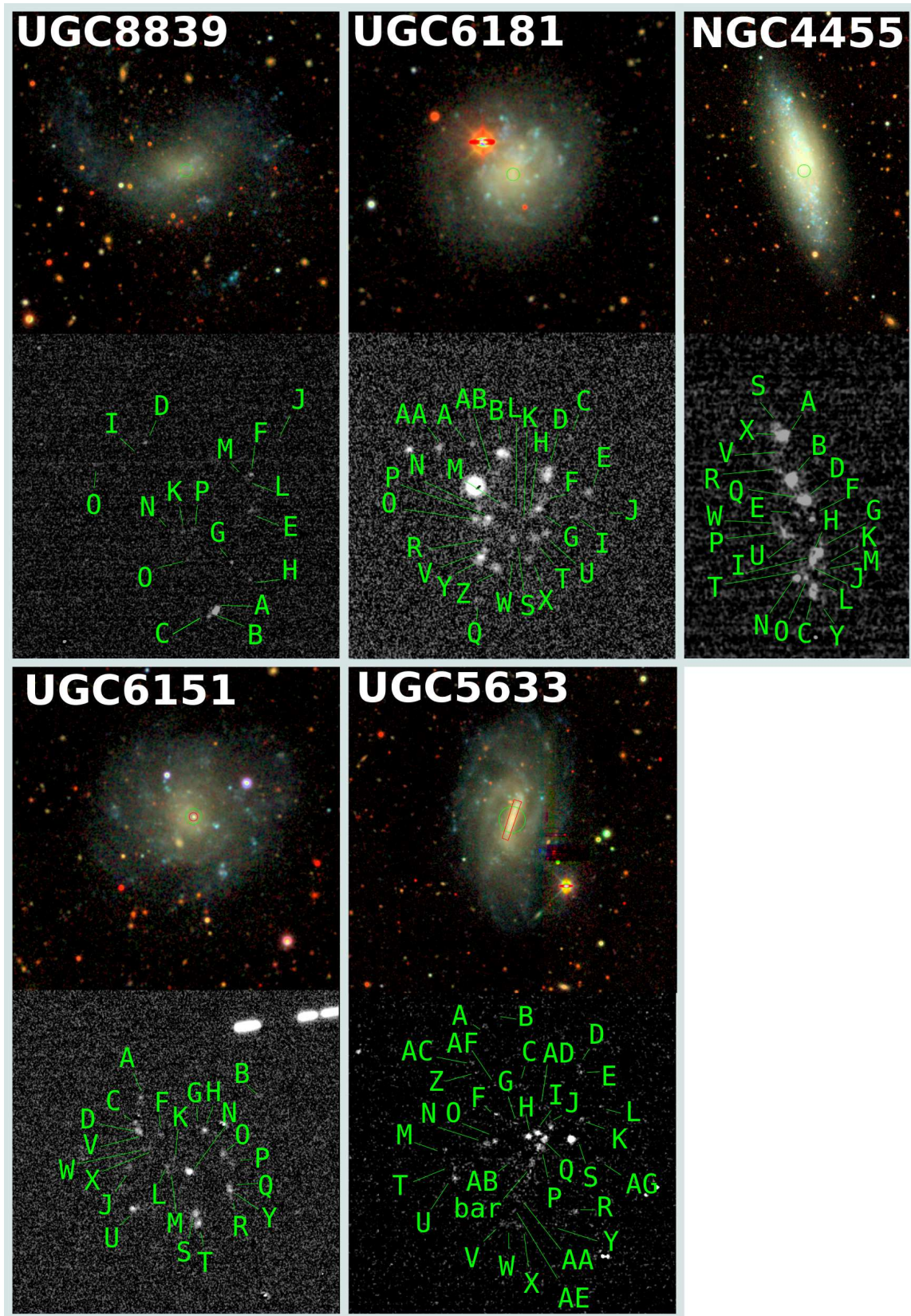}
  \caption{ Sample galaxies, DESI Legacy Imaging Survey grz images (top) and \HA{} images with \HII{} regions marked (bottom). The angular scale is the same in both sets of images. The \HA{} image of UGC~6151 has bright feature in the upper right corner; this is asteroid, and can be ignored.}
  \label{fig:images}
\end{figure*}

The calibrated images were visually inspected for \HII{} regions using SAOImage DS9, and cross-checked against publicly available sky surveys (SDSS DR16, 2MASS, DESI Legacy Imaging Survey). \HII{} regions were enclosed with a circular or elliptical boundary in SAOImage DS9 using the shape tool. Region clusters were separated into individual regions where definition between regions could be determined in both archival survey images and in our \HA{} images. These region boundaries were identified by careful inspection of the images to ensure that pixel brightness reached a minimum between regions. The \HA{} images, along with the identified \HII{} regions, are shown in \figref{images}. The fluxes within the regions are listed in \tblref{regions}. The uncertainties in the fluxes were derived by propagating in quadrature the calibration uncertainties and the standard deviation of the background (empty) pixels in the \HA{}+0nm and \HA{}+16nm images through the processes of continuum removal and the correction for the 2\AA{} equivalent width of stellar absorption underneath the \HA{} emission lines, as discussed in \secref{reduction}.

Additionally, whole-galaxy photometry was performed on the \HA{} images. All of the galaxies in our sample had at least some diffuse \HA{} emission not clearly associated with any individual region. The whole galaxy measurements and diffuse fractions are listed in \tblref{regions}.

It is worth noting that \cite{Richards2016} report a whole-galaxy \HA{} flux for UGC~8839 as $\log{f_\HA{}[\rm erg\,s^{-1}cm^{-2}]}=-13.61\pm0.18$, which differs from our value by more than $2\times$. This difference is driven by the fact that the \HA{} photometry in \cite{Richards2016} is confined to $R_{25}$ as measured in 3.6\micron{} Spitzer IRAC images. In the case of UGC~8839, $R_{25}=\arcsecond{34}{1}$. All of the \HII{} regions we report here are beyond this radius, and, as will be shown in subsequent sections, the \HA{} light in UGC~8839 is strongly biased toward the extreme outer disk.

\subsection{Broad-Band Surface Photometry}
\label{sec:gband}

In order to examine the significance of the location of the ABC complex on the outer edge of the disc of UGC~8839, \secref{location} discusses the local \band{broad} surface brightness around each \HII{} region within each of our sample galaxies. These surface brightness measurements were made using the publicly available DESI Legacy Imaging Survey \band{g,r,z} images and are listed in \tblref{regions}.

\begin{figure}
  \includegraphics[width=0.47\textwidth]{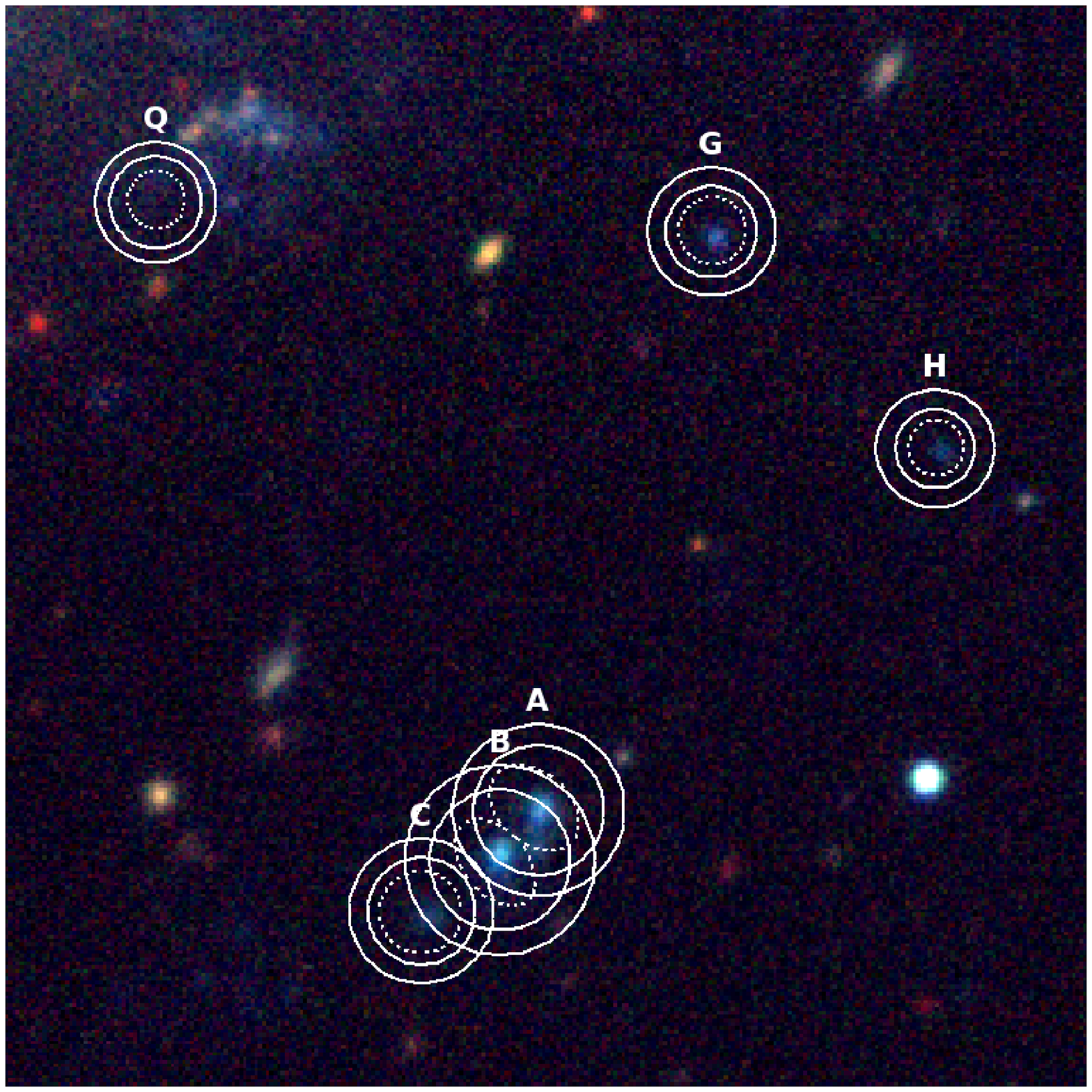}
  \caption{A close-up of the grz Legacy Sky Survey Image of UGC~8839 showing several \HII{} regions. The dashed shapes were used for photometry in the \HA{} images, and the solid annuli were used for surface photometry in the Legacy broad-band images.}
  \label{fig:annuli}
\end{figure}

Because the forthcoming analysis in \secref{location} relies on the local \band{g,r,z} surface brightness around each region being representative of the ``typical'' disc brightness at that galactocentric radius, it is necessary to exclude the \HII{} regions themselves from the local surface brightness calculation. After region identification (\secref{halpha}), a circular annulus with an inner radius of \arcsecond{1}{3} outside the \HII{} region and a width of \arcsecond{1}{3} was placed on each region. Adjustments by hand were made as necessary, and nearby clumps or local structures were masked out to ensure that the annulus sampled the true local disk. We also tuned the annular regions to be large enough that arm/inter-arm variations in surface brightness were averaged out. An example is shown in \figref{annuli}. All unmasked pixels in the broad-band images that were within each region's annulus and not within the photometry region of any other \HII{} region were averaged to generate a local surface brightness around each \HII{} region. This method is similar to the method used to determine the local surface brightness near \HII{} regions in \cite{Helmboldt2009}.

The central \band{g,r,z} surface brightness of each galaxy was measured as the average flux within a circular region placed at the visible centre (see green circular regions in \figref{images}). When possible, this circle was given a 5\arcsec{} radius. In NGC~4455 this was not possible because of the high density of bright knots (likely due to the high inclination of this galaxy). The bright \HII{} region near the centre of UGC~6151 and the bar in UGC~5633 also posed problems. Both of these bright objects were excluded (shown as red regions in \figref{images}). In UGC~5633, the green circle was expanded to 15\arcsec{}.

All surface brightnesses were corrected for foreground galactic extinction using $A_\lambda$ values listed on NED. We adopted the standard deviation of background (empty) pixels in the \band{g,r,z} images as uncertainties in the fluxes of individual galaxy pixels, and propagated those uncertainties forward through the surface brightness calculations. These uncertainties are listed in \tblref{regions}. Note that these uncertainties only account for random statistical uncertainties, and not systematic calibration uncertainties. Because the analysis in \secref{location} relies entirely on $\mu-\mu_0$, the effects of calibration uncertainties do not apply.

\section{Luminosity Distribution}
\label{sec:luminosity}

\begin{figure*}
  \includegraphics[height=0.85\textheight]{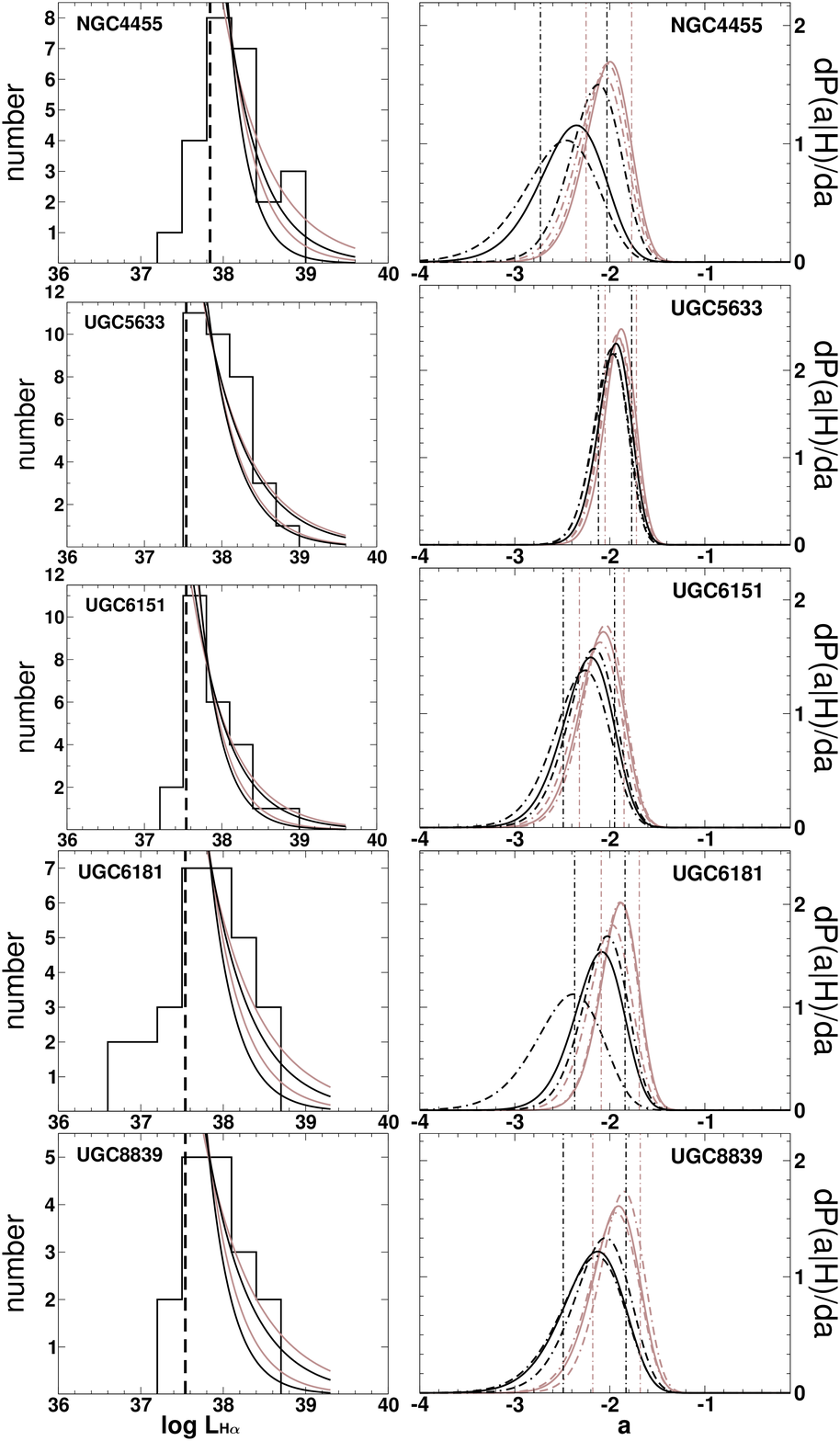}
  \caption{{\bf Left:} $\log{L_\HA{}}{\rm[erg\,s^{-1}]}$ histograms and smoothed density functions of the \HII{} regions in each galaxy. The vertical dashed line marks the peak of the smoothed density function, which we adopt as the completeness limit $L_C$. {\bf Right:} Probability density functions for the power-law index $a$. Although the histograms shown on the left are binned with $\Delta\log{L}{\rm[erg\,s^{-1}]}=0.3$, the probability density functions have been calculated for $\Delta\log{L}{\rm[erg\,s^{-1}]}=0.3$ (solid curve) and $\Delta\log{L}{\rm[erg\,s^{-1}]}=0.2,0.4$ (dashed curves). Note that they are not appreciably different, showing that the choice of bin width does not significantly impact our results. The $1\sigma$ limits on $a$ (for the $\Delta\log{L}{\rm[erg\,s^{-1}]}=0.3$ curves) are marked with dashed vertical lines in the right panels, and the expectation curves for those limits are superimposed on the histograms in the left panels as solid curves. The histograms generally follow the expectation curves. }
  \label{fig:histogram}
\end{figure*}

The left panels in \figref{histogram} show luminosity histograms for the \HII{} Regions in our sample galaxies. The histograms taper off slowly at high luminosities and abruptly at low luminosities. The upper end of the curve is likely real and representative of the HRLF, but the lower end of the curve is driven by incompleteness in our observations. The completeness limit, \lcomp{}, marked by the turnover point, is set by S/N in the images and also by the distances from Earth. \tblref{statistics} lists the measured \HA{} luminosity completeness limit (\lcomp{}) for each galaxy in our sample.

In order to test the hypothesis that the large \HII{} regions comprising the ABC complex in UGC~8839 are actually unusual, we will quantify the likelihood that UGC~8839 would host such regions by using the \HII{} region luminosity function (HRLF) as a likelihood predictor.

The primary challenge in fitting even a two parameter distribution to LSB galaxies is the scarcity of regions, often too few for standard ``best fit" methods given random variation. One solution is to stack the distributions from a sample of LSB galaxies and in \cite{Helmboldt2009}. This kind of technique is effective only in determining group-wide characteristics, and is not an ideal choice in this case since we are aiming to test the hypothesis that UGC~8839 has aberrant characteristics.

Instead, we organise the \HII{} regions of each galaxy into histogram bins, and then derive the probability $P(\geq\!N_{big}|H)$ that galaxy each will be able to populate its highest bin with $N_{big}$ regions given the distribution of the lower bins. We start in \secref{powerlaw} by calculating $\frac{dP(a|H)}{da}$, the probability density of the power-law index $a$ given that galaxy's observed histogram $H$ (excluding the highest bins). Then, in \secref{likelihood} we use $\frac{dP(a|H)}{da}$ to calculate $P(\geq\!N_{big}|H)$. If a galaxy's highest populated bin is an extreme outlier compared to the others, then $P(\geq\!N_{big}|H)$ will be very small, indicating that the most luminous regions are not drawn from the same population. A more moderate value of $P(\geq\!N_{big}|H)$, near 50\%, would indicate that the most luminous regions can be fully explained by the same HRLF that the smaller regions are drawn from.

\subsection{Determination of the Power Law Index}
\label{sec:powerlaw}

We start from the premise that the HRLF follows by a power law:

\begin{equation}
  \label{equ:hrlf}
  dN = A L^{a}dL
\end{equation}
where $A$ is a normalisation constant and $a$ is the power-law index. This power law must break down at low luminosities for physical reasons, however in practice the lower limit of its applicability is set by the completeness limit, $L_C$. For our analysis, we adopt the luminosity of the peak histogram bin as $L_C$.

We divide the luminosity range above $L_C$ into $B$ log luminosity bins (just as a histogram with $B$ bins). \figref{histogram} full shows histograms with a binning of $\Delta\log{L_\HA{}{\rm[erg\,s^{-1}]}}=0.3$, with $L_C$ marked as a vertical dashed line. We have repeated {the calculations which follow using $\Delta\log{L_\HA{}{\rm[erg\,s^{-1}]}}=0.2\text{\,and\,}0.4$. The results from all three sets of calculations are shown in \tblref{statistics}. Later in this section we demonstrate that the exact binning does not impact our general conclusions. If we consider \equref{hrlf} in the context of a single \HII{} region as a probability that that region will have a exact luminosity $L$, then the probability that that region will fall into bin $b$ is:

\begin{equation}
\label{equ:prob}
P_b = \int^{L_{b+1}}_{L_b} AL^{a}dL = \frac{A}{a+1}\left(L_{b+1}^{a+1}-L_b^{a+1}\right)
\end{equation}
where $A$ is defined to normalise the total probability.

Let $N$ be the total number of regions, and $n_b$ be the number of regions in bin $b$. Since each of the $n_b$ regions has the same expression for $P_b$, we can express the joint probability that all $n_b$ regions will fall within bin $b$ as:
\begin{equation}
P_b(n_b) = \binom{N}{n_b}\left(P_b\right)^{n_b} = \binom{N}{n_b}\pfrac{A}{a+1}^{n_b}\left(L_{b+1}^{a+1}-L_b^{a+1}\right)^{n_b}
\end{equation}

Note the presence of $\binom{N}{n_b}$ to account for number of combinations, which count identically. 

Finally, we multiply all the $P_b(n_b)$ together to find the joint probability that all the regions will fall within their respective bins to form the histogram $H$ given a value of the power-law index $a$: 
\begin{equation}
P(H|a) = \prod\limits_{b=0}^B P_b(n_b) = \prod\limits_{b=0}^B \binom{N_b}{n_b}\pfrac{A}{a+1}^{n_b}\left(L_{b+1}^{a+1}-L_b^{a+1}\right)^{n_b}
\end{equation}
Here $N_b=N-\sum\limits_{b'=0}^{b-1}${$n_b$}, the number of regions remaining after removing those falling into bins below bin $b$. Since the total number of regions $N = \sum\limits_{b=0}^{B}n_b$ we can write:

\begin{equation}
P(H|a) = \pfrac{A}{a+1}^N\prod\limits_{b=0}^B \binom{N_b}{n_b}\left(L_{b+1}^{a+1}-L_b^{a+1}\right)^{n_b}
\end{equation}

The expression above assumes a value for $a$, when in fact $a$ is uncertain. The differential probability of observing histogram $H$ is the joint probability of observing a histogram $H$ given $a$ and the differential probability observing the value of $a$:

\begin{equation}
dP(H|a) = \pfrac{A}{a+1}^N\prod\limits_{b=0}^B\binom{N_b}{n_b}\left(L_{b+1}^{a+1}-L_{b}^{a+1}\right)^{n_b}dP(a)
\end{equation}

Then, by Bayes' Theorem:
\begin{equation}
dP(a|H) = \frac{P(H|a)}{P(H)}dP(a)  = \frac{P(H|a)}{P(H)}\frac{dP(a)}{da}da
\end{equation}

To find $P(H)$, the total probability of observing the histogram $H$ regardless of the value of $a$, we integrate over all possible values of $a$:
\begin{equation}
P(H) = \int_{a_{min}}^{a_{max}}\frac{dP(H|a)}{da}da
\end{equation}

In our calculations we use the integral bounds $-4\!<\!a\!<\!-1$, however the exact choice of values for the integral bounds does not have an impact on the integral so long as $P(H|a)$ goes to zero for $a<-4$ and $a>-1$.  The final plots of $\frac{dP(a|H)}{da}$, shown in the right panels of \figref{histogram}, go to approximately zero for these limits, demonstrating that $P(H|a)$ must also go zero for these limits.

Likewise, the choice of prior is $\frac{dP(a)}{da}$ becomes moot for extreme values of $a$ since it is being multiplied by $P(H|a)$. For moderate values of $a$ the prior does have an impact. We assume a flat prior $\frac{dP(a)}{da}=1$; although a shallow power-law might be expected for late-type spirals, the HRLF is not well studied in LSB galaxies, and any strong assumptions on the prior would have a poor foundation.

Applying $P(H)$ and our flat prior yields a final expression for the probability density:
\begin{equation}
dP(H|a) = \frac{\pfrac{A}{a+1}^N\prod\limits_{b=0}^B\binom{N_b}{n_b}\left(L_{b+1}^{a+1}-L_{b}^{a+1}\right)^{n_b}da}{\int_{a_{min}}^{a_{max}}\pfrac{A}{a+1}^N\prod\limits_{b=0}^B\binom{N_b}{n_b}\left(L_{b+1}^{a'+1}-L_{b}^{a'+1}\right)^{n_b}da'}
\end{equation}

We have applied this expression to the histograms shown in the left panels in \figref{histogram} to determine the probability density for the HRLF power law index $a$, shown in the right panels in \figref{histogram}.

The solid curves show the probability density derived using bin widths of $\Delta\log{L_\HA{}}{\rm[erg\,s^{-1}]}=0.3$, the dashed curves using $\Delta\log{L_\HA{}}{\rm[erg\,s^{-1}]}=0.2$ and $0.4$. 
The black curves were generated excluding the highest populated bins, and the brown curves were generated using all the histogram bins. Note that the exclusion of the highest populated bin causes the curves to be broader (more uncertainty) and also skews the curves to steeper (lower) powers. The increase in uncertainty results from the diminished data set. The skew results from the fact that it is exclusively the high end data points being removed. In the analysis that follows we use the distribution with the highest luminosity bins removed (solid black curves) because our goal is to calculate the probability that each galaxy would host its largest regions  {\it given the distribution of the smaller regions}. That said, all the curves are generally similar to each other, indicating that  these assumptions do not have a significant impact on our analysis.

 Note also that these curves, which are essentially normalised $P(H|a)$, go to zero within the $-4\!<\!a\!<\!-1$ bounds, justifying our earlier assumptions about the range of $a$.

The dashed vertical lines in the right panels of \figref{histogram} show the upper and lower $1\sigma$ bounds for the $\Delta\log{L_\HA{}}{\rm[erg\,s^{-1}]}=0.3$ curves. The $1\sigma$ bounds and maximum probability values for all three probability curves are listed in \tblref{statistics}. Although the histograms in the left panels of \figref{histogram} are sparsely populated, all of the power-law indices are constrained to within $\pm0.4$ dex.

Using the $1\sigma$ limits as the upper and lower bounds on the power-law index, we have calculated upper and lower expectation curves for the number of \HII{} regions in each of the histogram bins, and over plotted those on top of the histograms as solid curves. Matching the colour scheme from the right panels, the black curves exclude the highest populated bins and the brown curves include them. Both sets of expectation curves match this histograms reasonably well, with random variation from small number counting noise. We conclude that the HRLF is well modelled by a power law based on the match between the expectation curves and the histograms.

\begin{table*}
\renewcommand{\arraystretch}{1.3}
\caption{HRLF Properties}
\begin{tabular}{lccccccccc}

&  &  & & & \multicolumn{2}{c}{\it largest included} & \multicolumn{2}{c}{\it largest excluded}\\
Galaxy & Bin Width & $L_C$  & $N\!>\!L_C$ & $N_{big}$ & $a$ & $P\left(N_{big}\right)$ &  $a$ &  $P\left(N_{big}\right)$\\
 & dex & erg s$^{-1}$ & & & \\
\hline
UGC8839	&	0.2	&	37.50	&	15	&	2	&	$-1.85^{+0.2}_{-0.3}$	&	0.58	&	$-2.04^{+0.3}_{-0.4}$	&	0.39	\\
UGC8839	&	0.3	&	37.50	&	15	&	2	&	$-1.91^{+0.3}_{-0.3}$	&	0.62	&	$-2.13^{+0.4}_{-0.4}$	&	0.41	\\
UGC8839	&	0.4	&	37.50	&	15	&	2	&	$-1.93^{+0.3}_{-0.3}$	&	0.70	&	$-2.14^{+0.4}_{-0.4}$	&	0.52	\\
NGC4455	&	0.2	&	37.80	&	20	&	1	&	$-2.02^{+0.3}_{-0.3}$	&	0.80	&	$-2.12^{+0.3}_{-0.3}$	&	0.73	\\
NGC4455	&	0.3	&	37.80	&	20	&	3	&	$-2.00^{+0.3}_{-0.3}$	&	0.45	&	$-2.35^{+0.4}_{-0.5}$	&	0.18	\\
NGC4455	&	0.4	&	37.80	&	20	&	3	&	$-2.06^{+0.3}_{-0.3}$	&	0.51	&	$-2.46^{+0.4}_{-0.5}$	&	0.22	\\
UGC6181	&	0.2	&	37.50	&	22	&	2	&	$-1.89^{+0.2}_{-0.3}$	&	0.72	&	$-2.03^{+0.3}_{-0.3}$	&	0.57	\\
UGC6181	&	0.3	&	37.50	&	22	&	3	&	$-1.88^{+0.2}_{-0.3}$	&	0.64	&	$-2.08^{+0.3}_{-0.3}$	&	0.40	\\
UGC6181	&	0.4	&	37.50	&	22	&	4	&	$-1.97^{+0.2}_{-0.3}$	&	0.47	&	$-2.38^{+0.4}_{-0.5}$	&	0.16	\\
UGC6151	&	0.2	&	37.50	&	23	&	1	&	$-2.05^{+0.3}_{-0.3}$	&	0.68	&	$-2.16^{+0.3}_{-0.3}$	&	0.57	\\
UGC6151	&	0.3	&	37.50	&	23	&	1	&	$-2.07^{+0.3}_{-0.3}$	&	0.66	&	$-2.20^{+0.3}_{-0.4}$	&	0.54	\\
UGC6151	&	0.4	&	37.50	&	23	&	1	&	$-2.10^{+0.3}_{-0.3}$	&	0.62	&	$-2.26^{+0.3}_{-0.4}$	&	0.49	\\
UGC5633	&	0.2	&	37.50	&	33	&	1	&	$-1.92^{+0.2}_{-0.2}$	&	0.89	&	$-1.98^{+0.2}_{-0.2}$	&	0.85	\\
UGC5633	&	0.3	&	37.50	&	33	&	1	&	$-1.88^{+0.2}_{-0.2}$	&	0.91	&	$-1.93^{+0.2}_{-0.2}$	&	0.88	\\
UGC5633	&	0.4	&	37.50	&	33	&	1	&	$-1.91^{+0.2}_{-0.2}$	&	0.90	&	$-1.97^{+0.2}_{-0.2}$	&	0.85	\\

\label{tbl:statistics}
\end{tabular}
\end{table*}

\subsection{Likelihood of 1st Ranked Bin}
\label{sec:likelihood}

In order to determine whether the ABC complex is extraordinary given the entire population of \HII{} regions in UGC~8839, we will derive a general formula for the probability  $P(\geq\!N_{big})$ that a galaxy with $N$ \HII{} regions and a probability density function $\frac{dP(a|H)}{da}$ will have at least $N_{big}$ large regions, where ``large'' is defined as the most luminous populated histogram bin. This probability is a measure of the HRLF's ability to explain the presence of the most luminous \HII{} regions. If a galaxy's most luminous populated bin is an extreme outlier then $P(\geq\!N_{big})$ will be very small, indicating that it is very \emph{unlikely} that the most luminous regions in that galaxy are drawn from the same population as the less luminous ones. 

As long as $a\!<\!-1$, we can take the upper limit in \equref{prob} to infinity to find the total probability of a given \HII{} region having a luminosity greater than $L$:

\begin{equation}
  P(\geq\!L) = -\frac{A L^{a+1}}{a+1}
  \label{equ:nfunc}
\end{equation}

Note that this expression is positive despite the negative sign since $a+1<0$.

Since we are considering only \HII{} regions with $L_\HA{}>\lcomp{}$, we  can normalise to $P(\geq\! L_\lcomp{})=1$, and let $L=L_{big}$, the threshold luminosity for an \HII{} region to be included in the largest histogram bin:

\begin{equation}
  P(\geq\!L_{big}|a) = \pfrac{L_{big}}{\lcomp{}}^{a+1}
\end{equation}

Given this expression for the probability of an individual \HII{} region having $L_\HA{}>L_{big}$, we now consider a galaxy with $N$ \HII{} regions. The joint probability of {\it exactly} $N_{big}$  regions all having luminosities $L_\HA{}>L_{big}$ can be expressed as the joint probability that $N_{big}$ regions will have $L_\HA{}>L_{big}$ and $N-N_{big}$ regions will have $L_\HA{}<L_{big}$, multiplied by $\binom{N}{N_{big}}$, the number of combinations of $N_{big}$ regions drawn from a sample of $N$:

\begin{equation}
  P(N_{big}|a) = \binom{N}{N_{big}}\pfrac{L_{big}}{\lcomp{}}^{(a+1)N_{big}} \left(1-\pfrac{L_{big}}{\lcomp{}}^{a+1}\right)^{N-N_{big}} 
  \label{equ:P_N}
\end{equation}

Note that if we sum $P(N_{big})$ over $N_{big}\!\!=\!\!0$ to $N_{big}\!\!=\!\!N$, the sum is unity.

However, this expression for $P(N_{big}|a)$ is contingent upon the value of the power-law index $a$, which is itself uncertain. We can express this by using \equref{P_N} to find the joint probability of a power-law index with a value of exactly $a$ and, given that value, also having exactly $N_{big}$ regions with luminosities $L_\HA{}>L_{big}$:

\begin{equation}
P(N_{big}) = P(N_{big}|a)P(a)
\end{equation}

As discussed in \secref{powerlaw}, the number of \HII{} regions in our sample galaxies is too few for standard ``best fit" methods given random variation, so instead we derived a probability density function $\frac{dP(a|H)}{da}$ given the observed histogram $H$. Using that result, Our expression for $P(N_{big})$ then becomes a differential probability

\begin{equation}
dP(N_{big}) = P(N_{big}|a)dP(a|H) = P(N_{big}|a)\frac{dP(a|H)}{da}da
\end{equation}

Which then becomes an integral:

\begin{equation}
  P\left(N_{big}\right)   = \int_{a_{min}}^{a_{max}}P(N_{big},a)\, \frac{dP(a|H)}{da}\,da
  \label{equ:dP_N}
\end{equation}

This is is the total probability that, given the histogram $H$, a galaxy will have exactly $N_{big}$ regions in its 1st ranked bin. However, in our case, we are interested in testing the hypothesis that the 1st ranked bin is anomalously overpopulated. To do this, we will calculate the probability that the 1st ranked bin will be populated with at least the number of regions as in our measurements:

\begin{equation*}
P\left(\geq\!N_{big}\right) = 1-\sum\limits_{N_{big}'=0}^{N_{big}} P(N_{big}')
\end{equation*}

Using this formula,  we present in \tblref{statistics} the calculated values of $P\left(\geq\!N_{big}\right)$, the probabilities that the 1st ranked luminosity bins would be populated with at least $N_{big}$ regions given the total number of regions $N$ with $L_\HA{}\!>\!L_C$ and the probability density $\frac{dP(a|H)}{da}$. These results will be discussed further in \secref{discussion}, however we find that the probability that UGC~8839 will be able to populate its 1st ranked bin with the observed $N_{big}$ is 40-50\% for all choices of bin width. We conclude that presence of the ABC complex is not extraordinary given entire observed population of \HII{} regions in UGC~8839, and that a single power-law HRLF is fully capable of explaining presence of the ABC complex in UGC~8839. The implications of this conclusion will be discussed in greater depth in \secref{discussion}, however the main point is that, although the ABC complex visually stands out and seems ``out of place'' in UGC~8839 (\figref{contours}), it is actually in-line with the entire population of \HII{} regions.

\section{Spatial Distribution}
\label{sec:location}

In \secref{luminosity} we showed that the presence of the ABC complex in UGC~8839 is not extraordinary. The other striking aspect of the ABC complex is its location: this complex, which accounts for 69\% of the discrete \HA{} light, is located 8~kpc from the galaxy's centre, beyond the main body of  the visible disc (\figref{images}). To explore the seemingly odd placement for such a large region in an otherwise quiet galaxy, we quantify the spatial distribution of star formation in the remaining galaxies in our sample and compare them to UGC~8839.

When comparing the spatial distribution of \HII{} regions in different galaxies, it is essential that we compare them on equal footing. For example, the ABC complex's location of 8~kpc from the centre of UGC~8839 puts it in the extreme outer disk of this small galaxy. In comparison, 8~kpc is roughly the solar circle in the larger Milky Way.

Instead of comparing the spatial distribution in terms in physical distances, it is more sensible to compare them using a characteristic distance. Since these galaxies are all late-types, the exponential scale length is a natural length to use, defined by:

\begin{equation}
\mu=\mu_0 + \frac{2.5}{\ln{10}}\frac{r}{\alpha}
\label{equ:exponential}
\end{equation}

Here $r$ is the deprojected angular distance from the centre of the galaxy, $\alpha$ is the angular exponential scale length of the galaxy, and $\mu$ and $\mu_0$ are the local and central surface brightnesses in \surf{}. Expressing the position of the \HII{} regions within their respective galaxies in terms of $\frac{r}{\alpha}$  rescales the galactocentric distances to the size of each galaxy. Also, since both measurements are in angular units, any uncertainties or systematic effects in the heliocentric distances are excluded. Note that many galaxies are known to have piece-wise light profiles or ``broken exponential'' profiles, and we caution against over interpretation of $\frac{r}{\alpha}$ as a literal number of scale lengths. Instead, we are using it here as a characteristic distance.

However, the practical application of \equref{exponential} requires the fitting of the disc profile, which would inject a significant amount of uncertainty given the faintness of our galaxies. For example, the deprojected angular positions of the \HII{} regions ($r$) would be heavily influenced by uncertainties in the inclination angle and the position angle of their host galaixes. Even discounting these issues, a straightforward application of \equref{exponential} would be cumbersome for galaxies with broken exponential profiles.

Instead, we use $\left(\mu-\mu_0\right)\frac{\ln{10}}{2.5}$, which is equivalent to $\frac{r}{\alpha}$ but sidesteps uncertainties about disc geometry. As discussed in \secref{gband}, $\mu$ was measured locally around each \HII{} region, and $\mu_0$ was measured in the geometric centre of each galaxy. Because both measurements are local, no corrections for inclination or geometry are needed. Also, because an annulus was used to measure the local surface brightness around each region, the $\mu$ values for each region do not reflect the characteristics of that region, but instead reflect the characteristics of the disc at each galactocentric radius.

Azimuthal variations in $\mu$ from spiral arms as well as general clumpiness pose a concern with our use of $\mu-\mu_0$ as a proxy for $\frac{r}{\alpha}$. As discussed  in \secref{gband}, the annuli were chosen in such a way to mitigate these issues. Additionally, our analysis uses three filters, g, r, and z, which provides a secondary check: any effect seen only or primarily in the \band{g} would be suspect on the grounds that galaxies are clumpier and have higher arm/inter-arm contrast in bluer bands. Conversely, any conclusions drawn based on all three bands, especially the redder bands, are unlikely to be due to arms or clumpiness associated with star formation.

\begin{figure*}
  \includegraphics[width=0.9\textwidth]{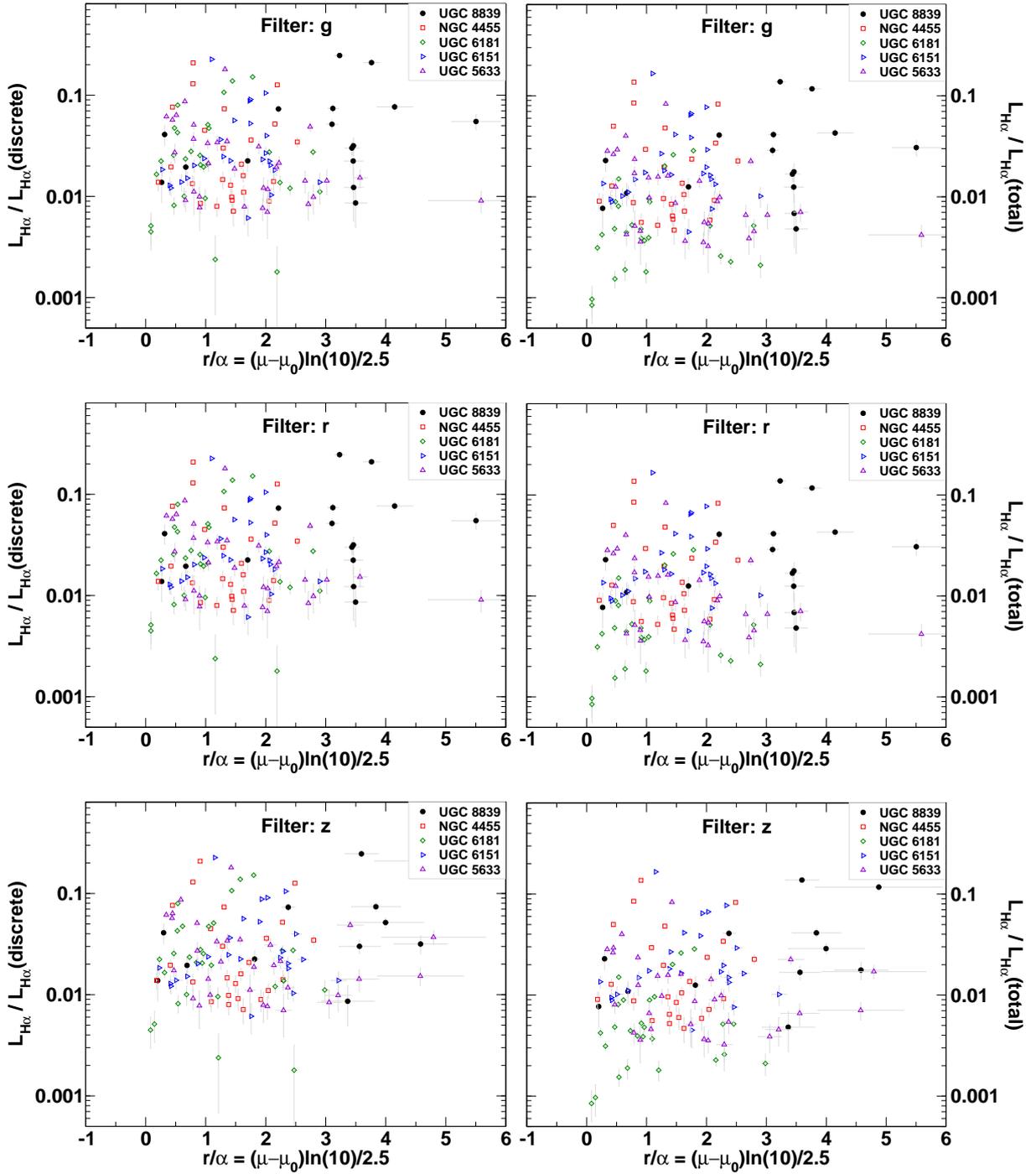}
  \caption{ All the \HII{} regions in our study, plotted as fractions of total \HA{} luminosity versus  relative distance from the galaxy centre. The plots on the left calculate the total luminosity as the sum of the discrete \HII{} regions, while the plots on the right include the diffuse emission. In both cases and in all filters the regions from UGC~8839 fall on the upper right envelope of the plot. Star formation in UGC~8839 shows a clear preference for the disc edge.}
  \label{fig:location}
\end{figure*}

\figref{location} shows all the \HII{} regions in our sample plotted by their fractional contribution to the total \HA{} light versus $\left(\mu-\mu_0\right)\frac{\ln{10}}{2.5}$ for the g, r, and z filters. In the left panels, the total \HA{} light is the sum of all the discrete \HII{} regions. In the panels on the right, the total \HA{} light includes the diffuse emission. Note that the overall pattern of points in the left and right panels is not appreciably different.

In all panels most of the \HII{} regions fall within a general cloud which is not significantly different from galaxy to galaxy, with the notable exception of UGC~8839. The regions associated with UGC~8839 lie primarily on the upper right envelope of the distribution. For their luminosities, the \HII{} regions in UGC~8839 are almost all located farther from the centre than analogous region in the remaining galaxies. The remaining galaxies show no discernible patterns. We find that star formation in UGC~8839 shows a marked proclivity for the extreme outer disk. The implications of disc-edge dominant star formation in UGC~8839 will be discussed further in \secref{hyperstable}.

Note that the main effect, that the \HII{} regions in UGC~8839 are preferentially on the disk edge (at larger $\frac{r}{\alpha}$), is strongest in the \band{r} and still very strong in the \band{z}, where the galaxies are the smoothest and the least clumpy. This shows that that asymmetric features such as a arms and clumps do not contribute significantly to our conclusions. 

The consistency of this effect across filters also addresses another potential pitfall: that errors in the measurement of $\mu_0$ would shift all the points for a given galaxy left or right. Since the measurements in the different filters are independent, we can eliminate this possibility. Additionally, the difference between the points for UGC~8839 and the remaining galaxies is approximately one half to one magnitude, far larger than our measurement errors in $\mu_0$.

\section{Discussion}
\label{sec:discussion}

\subsection{General Comments on the HRLF}
\label{sec:general_comments}

The power law indices in table \tblref{statistics} are all around $-2$. This consistent with expectations, since most late-type HSB galaxies are best characterised by power laws in the range $-2\!<\!a\!<\!-1.5$ \citep[e.g.,][]{KEH1989}.

The fraction of \HA{} light diffusely associated with each galaxy and not with identifiable regions spans a  range, from 27\% in UGC~6151 to 81\% in UGC~6181. All except UGC~6151 and the HSB NGC~4455 have diffuse fractions greater than 40\%, generally in agreement the findings in \cite{ONeil2007}, which show that LSB galaxies tend to have higher diffuse fractions of \HA{} light than HSB galaxies, usually above 50\%. It is not possible for us to determine what fraction of the diffuse light may be unresolved \HII{} regions below our completeness limit $L_C$, however the the consistent power law of $-2$ despite a $>\!3\times$ variation in the diffuse fraction suggests that this effect is not significantly impacting our results. The southern end of the bar in UGC~5633 is even evident in the \HA{} images as diffuse emission, although it only contributes about 5\% to the total diffuse emission.

\subsection{The ABC Complex as an Intruder}
\label{sec:intruder}

One possible explanation for the presence of the ABC complex in the extreme outer disc UGC~8839 is that the complex is an intruder object, perhaps a small gas-rich dwarf merging with UGC~8839. For the following reasons, the analysis above shows that this interpretation is unlikely:

First, in \secref{luminosity} we examined the probability that UGC~8839 would host its largest regions given the population of smaller \HII{} regions, and found that the probability is $\gtrsim\!40\%$. While the uncertainties due to low number statistics are significant, we can safely conclude that the existence of the ABC complex in UGC~8839 is not, by itself, extraordinary. The HRLF of UGC~8839 is fully capable of accounting for the ABC complex, without requiring that the complex be an intruder object.

Second, in \secref{location} we examined the location of \HII{} regions in our sample galaxies by plotting the fractional contribution of each region to its host galaxy's total \HA{} light against the local \band{broad} surface brightness in the immediate vicinity of that \HII{} region (excluding the region itself). We found that the \HII{} regions in UGC~8839 fall on the upper right envelope of the distribution; for their luminosities, they are, on average, farther out in the disc of UGC~8839 than comparable regions in our other sample galaxies.

Although this investigation was motivated by the striking appearance of the ABC complex, we find that the entire population of \HII{} regions in UGC~8839 is unusual. Given that the ABC complex does not, in fact, stand out as unusual when taken in the context of the smaller \HII{} regions in UGC~8839, we conclude there is no compelling evidence that the ABC complex is an intruder object, and that it can be fully explained within the context of the \HII{} region population in UGC~8839.

\subsection{UGC~8839 as a Mini-Malin 1}
\label{sec:minimalin}

Alternatively, if indeed UGC~8839 has experienced a recent merger with a smaller gas-rich satellite, the intruder may have been absorbed or disrupted, and its gas is now fuelling star formation within the disc of UGC~8839. If this kind of event is responsible for most or all of the \HII{} regions in UGC~8839, then we would not expect the ABC complex to stand out as distinct.

This hypothesis has precedent: In the last decade a number of works have shown that gas-rich mergers are able to stimulate starbursts in small galaxies just as they are in more massive galaxies. As a somewhat extreme but well-studied example, \cite{Ekta2008} report a disturbed \HI{} morphology in the blue compact dwarf galaxy DDO~68, suggesting that the current star formation is merger driven; this was followed by several detailed studies of the stellar population and ISM metallicity \citep{MartinezDelgado2012,Annibali2016}, which confirm that DD0~68 is a late-stage merger between the main progenitor of DDO~68 and a smaller gas-rich satellite. Although DDO~68 is much smaller than UGC~8839 ($M_g=-16.2$ mag. versus $M_g=-17.3$ mag.), they do have similar gas-mass to light ratios: $\mathcal{M}_{\rm HI}/L_g=2.4$ for DDO~68 \citep{Ekta2008} and $\mathcal{M}_{\rm HI}/L_g=2.6$ for UGC~8838 \citep{Richards2016}  (solar units).

However, the \HI{} map of UGC~8839 shown in Figure A15 of \cite{Richards2016} shows that, although the \HI{} disc does extend out as far as the ABC complex, it shows no signs of disrupted morphology, unlike the disturbed \HI{} morphology in DDO~68.

Additionally, the \HI{} gas and the star formation in DDO~68 are located near the galactic centre \citep{Pustilnik2005,Ekta2008}, consistent with the angular moment loss associated with a typical accretion event. In practice, it would be difficult to arrange the merger and disruption of a gas-rich dwarf without a significant loss of angular momentum and disrupted morphology, which is not seen in UGC~8839.

A three-body interaction may be the solution to this puzzle. The giant LSB galaxy Malin~1, originally thought to be an early-type galaxy, is now known to have a very faint, blue, gas-rich disc with an incredible scale length of $\sim$200kpc. The origin of the faint blue disc remained a puzzle until \cite{Zhu2018} reported a Malin~1 analogue in the IllustrisTNG simulation. Looking backward in the simulation, they find that the cause of the extended gas disc is a three-body interaction, wherein an interacting pair of gas-rich galaxies merge with a larger gas-poor object.  This three-body interaction allows the larger galaxy to acquire the satellites' gas without significantly heating it or driving it to the centre of the newly merged core, forming a diffuse extended disk. It may be that UGC~8839 is a miniature analogue to Malin~1.

In this scenario, a three-body interaction delivered gas to UGC~8839, creating the outer disc, and the merging satellites have since been absorbed or disrupted. As with the proposed origin for Malin~1, the particular details of the three-body interaction would lead to the gas being removed from the satellites while retaining enough angular momentum to form the extended gas disc of UGC~8839. Our findings here, that star formation in UGC~8839 primarily occurs in the extreme outer disk, are consistent with this scenario.

\subsection{UGC 8839 as a Hyper-Stable Disc}
\label{sec:hyperstable}

Although a three-body interaction may explain the origin of the extended disc and the ABC complex, further explanation is required for UGC~8839 as a whole. The \HI{} maps reported in \cite{Richards2016} show that the \HI{} is more concentrated near the centre of UGC~8839 than near the ABC complex. Although the ABC complex is located at a local \HI{} density maximum, that maximum is less dense than the \HI{} at the centre of the galaxy where we see very little \HA{} emission. The most curious aspect of UGC~8839 is not that it supports star formation, but rather that the star formation does not seem to follow the gas.

A plausible explanation is put forth in \cite{Garg2017}, wherein the authors calculate the modified Toomre parameter developed in \cite{Romeo2011} for a sample of LSB discs. This modified Toomre parameter accounts for gas as well as stars; since LSB discs are typically more gas rich than HSB discs, they consider the possibility that too much gas may `over stabilise' a disc against star formation. However, \cite{Garg2017} find that the most significant impact on LSB disc stability comes from the fact they tend to be dark matter dominated at all radii \citep{deBlok1997,Simon2003,Banerjee2010,KuziodeNaray2011}. As a result, several of the galaxies in their sample reach stability minima only in the outer disc, several scale lengths from the centre.

The dynamical analysis based on the \HI{} velocity field data presented in \cite{Richards2016} confirms that, for reasonable assumptions about baryonic $\mathcal{M}/L$, UGC~8839 is dark matter dominated all the way to its centre. Indeed, UGC~8839 is overall the most dark matter dominated galaxy in the sample of 15 galaxies presented in \cite{Richards2016}.

We find this to be the most likely interpretation, that a higher dark matter fraction and greater dark matter dominance causes the disc of UGC~8839 to be more stable than the other galaxies in our sample. This causes the local stability minimum to be located beyond the edge of the visible disc, and explains the preference of star formation for the outer disc of seen in \figref{location}. The ABC complex is explained as a structure native to UGC~8839, which is consistent with the statistical analysis of the HRLF presented in \secref{luminosity}. The extreme edge-dominated nature of star formation in UGC~8839, of which the ABC complex is simply the most visible example, is indicative of galactic dynamics dominated by dark matter at all radii.

\section{Summary}
\label{sec:summary}

We examined new \HA{} images and archival DESI Legacy Imaging Survey \band{g,r,z} images of four LSB spirals and one HSB spiral. We used this information to assess the significance of the large \HII{} region complex in the extreme outer disc of the LSB spiral UGC~8839 (the ABC complex), and then used this information to discuss the origins and nature of the ABC complex.

Adopting a power law \HII{} region luminosity function, we determined the probability density function for the power law index for each galaxy. Using these probability density curves, we then calculated the likelihood that each galaxy would host its most luminous \HII{} regions given a random sample equal to the number of actual regions in that galaxy. We find that the likelihood of the ABC complex being present in UGC~8839 given the distribution of smaller regions is moderate, 40\%-50\%, and conclude that the presence of the ABC complex in UGC~8839 is not extraordinary since it can be fully accounted as part of the ensemble of \HII{} regions.

We then examined the spatial distribution of \HII{} regions in all of our sample galaxies, and find that the \HII{} regions in UGC~8839 are preferentially found multiple scale lengths from the centre, whereas the other galaxies in our sample have a more centrally concentrated distribution. We conclude that the spatial distribution of \HII{} regions in UGC~8839 is unusual, and that the ABC complex is simply the most visible example.

The most likely interpretation of these findings is that the disc of UGC~8839 is extremely stable against cloud formation because of a high dark matter fraction \citep{Richards2016}. In particular, UGC~8839 is likely dark matter dominated even in its central regions, which would raise the Toomre and modified Toomre parameters ($Q$,$Q_{RW}$) to levels that inhibit star formation and push the stability minimum past the edge of the visible galactic disc.

This highlights the fundamental link between the dark matter halos and star-formation histories of galaxies. UGC~8839 is a fairly extreme example in both categories, and hints at the possibility that galaxies with greater dark/baryonic matter fractions might show an even greater proclivity for edge-dominated star formation.

A three-body interaction, similar to the interaction suspected to be responsible for the extended gas disc in the giant LSB galaxy Malin~1 \citep{Zhu2018},  may be able to explain the extended gas disc in UGC~8839. However, a three-body interaction would not, by itself, be able to account for the anti-correlation of \HI{} density and star formation, wherein the core of UGC~8839 has a high \HI{} density but very little star formation.

\section*{Acknowledgements}

We gratefully acknowledge the financial support of Five Colleges, Inc. and the technical support of the staff of the WIYN 0.9m telescope at KPNO.

Based on observations at Kitt Peak National Observatory, NSF’s National Optical-Infrared Astronomy Research Laboratory, which is operated by the Association of Universities for Research in Astronomy (AURA) under a cooperative agreement with the National Science Foundation.

Funding for the Sloan Digital Sky Survey IV has been provided by the Alfred P. Sloan Foundation, the U.S. Department of Energy Office of Science, and the Participating Institutions. SDSS-IV acknowledges
support and resources from the Center for High-Performance Computing at
the University of Utah. The SDSS web site is www.sdss.org.

SDSS-IV is managed by the Astrophysical Research Consortium for the  Participating Institutions of the SDSS Collaboration including the Brazilian Participation Group, the Carnegie Institution for Science, Carnegie Mellon University, the Chilean Participation Group, the French Participation Group, Harvard-Smithsonian Center for Astrophysics,  Instituto de Astrof\'isica de Canarias, The Johns Hopkins University, Kavli Institute for the Physics and Mathematics of the Universe (IPMU) / University of Tokyo, the Korean Participation Group, Lawrence Berkeley National Laboratory, Leibniz Institut f\"ur Astrophysik Potsdam (AIP),  Max-Planck-Institut f\"ur Astronomie (MPIA Heidelberg), Max-Planck-Institut f\"ur Astrophysik (MPA Garching), Max-Planck-Institut f\"ur Extraterrestrische Physik (MPE), National Astronomical Observatories of China, New Mexico State University, New York University, University of Notre Dame, Observat\'ario Nacional / MCTI, The Ohio State University, Pennsylvania State University, Shanghai Astronomical Observatory, United Kingdom Participation Group, Universidad Nacional Aut\'onoma de M\'exico, University of Arizona, 
University of Colorado Boulder, University of Oxford, University of Portsmouth, University of Utah, University of Virginia, University of Washington, University of Wisconsin, Vanderbilt University, and Yale University.

The Legacy Surveys consist of three individual and complementary projects: the Dark Energy Camera Legacy Survey (DECaLS; NOAO Proposal ID \# 2014B-0404; PIs: David Schlegel and Arjun Dey), the Beijing-Arizona Sky Survey (BASS; NOAO Proposal ID \# 2015A-0801; PIs: Zhou Xu and Xiaohui Fan), and the Mayall z-band Legacy Survey (MzLS; NOAO Proposal ID \# 2016A-0453; PI: Arjun Dey). DECaLS, BASS and MzLS together include data obtained, respectively, at the Blanco telescope, Cerro Tololo Inter-American Observatory, National Optical Astronomy Observatory (NOAO); the Bok telescope, Steward Observatory, University of Arizona; and the Mayall telescope, Kitt Peak National Observatory, NOAO. The Legacy Surveys project is honored to be permitted to conduct astronomical research on Iolkam Du'ag (Kitt Peak), a mountain with particular significance to the Tohono O'odham Nation.

NOAO is operated by the Association of Universities for Research in Astronomy (AURA) under a cooperative agreement with the National Science Foundation.

This project used data obtained with the Dark Energy Camera (DECam), which was constructed by the Dark Energy Survey (DES) collaboration. Funding for the DES Projects has been provided by the U.S. Department of Energy, the U.S. National Science Foundation, the Ministry of Science and Education of Spain, the Science and Technology Facilities Council of the United Kingdom, the Higher Education Funding Council for England, the National Center for Supercomputing Applications at the University of Illinois at Urbana-Champaign, the Kavli Institute of Cosmological Physics at the University of Chicago, Center for Cosmology and Astro-Particle Physics at the Ohio State University, the Mitchell Institute for Fundamental Physics and Astronomy at Texas A\&M University, Financiadora de Estudos e Projetos, Fundacao Carlos Chagas Filho de Amparo, Financiadora de Estudos e Projetos, Fundacao Carlos Chagas Filho de Amparo a Pesquisa do Estado do Rio de Janeiro, Conselho Nacional de Desenvolvimento Cientifico e Tecnologico and the Ministerio da Ciencia, Tecnologia e Inovacao, the Deutsche Forschungsgemeinschaft and the Collaborating Institutions in the Dark Energy Survey. The Collaborating Institutions are Argonne National Laboratory, the University of California at Santa Cruz, the University of Cambridge, Centro de Investigaciones Energeticas, Medioambientales y Tecnologicas-Madrid, the University of Chicago, University College London, the DES-Brazil Consortium, the University of Edinburgh, the Eidgenossische Technische Hochschule (ETH) Zurich, Fermi National Accelerator Laboratory, the University of Illinois at Urbana-Champaign, the Institut de Ciencies de l'Espai (IEEC/CSIC), the Institut de Fisica d'Altes Energies, Lawrence Berkeley National Laboratory, the Ludwig-Maximilians Universitat Munchen and the associated Excellence Cluster Universe, the University of Michigan, the National Optical Astronomy Observatory, the University of Nottingham, the Ohio State University, the University of Pennsylvania, the University of Portsmouth, SLAC National Accelerator Laboratory, Stanford University, the University of Sussex, and Texas A\&M University.

BASS is a key project of the Telescope Access Program (TAP), which has been funded by the National Astronomical Observatories of China, the Chinese Academy of Sciences (the Strategic Priority Research Program ``The Emergence of Cosmological Structures'' Grant \# XDB09000000), and the Special Fund for Astronomy from the Ministry of Finance. The BASS is also supported by the External Cooperation Program of Chinese Academy of Sciences (Grant \# 114A11KYSB20160057), and Chinese National Natural Science Foundation (Grant \# 11433005).

The Legacy Survey team makes use of data products from the Near-Earth Object Wide-field Infrared Survey Explorer (NEOWISE), which is a project of the Jet Propulsion Laboratory/California Institute of Technology. NEOWISE is funded by the National Aeronautics and Space Administration.

The Legacy Surveys imaging of the DESI footprint is supported by the Director, Office of Science, Office of High Energy Physics of the U.S. Department of Energy under Contract No. DE-AC02-05CH1123, by the National Energy Research Scientific Computing Center, a DOE Office of Science User Facility under the same contract; and by the U.S. National Science Foundation, Division of Astronomical Sciences under Contract No. AST-0950945 to NOAO.

\section*{Data Availability}

The \HA{} data underlying this article will be shared on reasonable request to the corresponding author, and are also available via the NOAO Science Archive:

http://archive1.dm.noao.edu/search/query/

The SDSS data are publicly available via the SDSS sky server:

http://skyserver.sdss.org/dr16/en/tools/chart/navi.aspx

\bibliographystyle{mnras}
\bibliography{lsb}



\appendix

\section{Data on Individual Regions}
\pagebreak

\begin{table*}
\caption{Properties of Individual \HII{} Regions}
\label{tbl:regions}
\begin{tabular}{ccccccccccc}
	& $f_\HA{}$ & $\log{L_\HA{}}$ & $\Delta\log{L_\HA{}}$ & $\frac{f_\HA{}}{f_\HA{}(tot)}$ & $\mu(g)$ & $\Delta\mu(g)$   & $\mu(r)$ & $\Delta\mu(r)$   & $\mu(z)$ & $\Delta\mu(z)$	\\
    & $\rm10^{-15}erg\,s^{-1}\,cm^{-2}$ & $\log{\rm erg\,s^{-1}}$ & & & \surf{}& \surf{}& \surf{}& \surf{}& \surf{}& \surf{} \\
 \hline\\	
\underline{UGC8839}	&&&\\														
A	&	6.81	&	38.60	&	0.01	&	0.117	&	27.0	&	0.13	&	26.9	&	0.20	&	27.6	&	1.06	\\
B	&	8.01	&	38.67	&	0.01	&	0.138	&	26.5	&	0.08	&	26.4	&	0.13	&	26.2	&	0.28	\\
C	&	2.40	&	38.15	&	0.04	&	0.041	&	26.3	&	0.09	&	26.1	&	0.11	&	26.5	&	0.41	\\
D	&	1.79	&	38.02	&	0.08	&	0.031	&	28.9	&	0.41	&	28.3	&	0.40	&	---	&	---	\\
E	&	2.37	&	38.15	&	0.04	&	0.041	&	25.4	&	0.04	&	25.2	&	0.06	&	24.9	&	0.11	\\
F	&	2.49	&	38.17	&	0.04	&	0.043	&	27.5	&	0.30	&	27.1	&	0.33	&	---	&	---	\\
G	&	1.68	&	38.00	&	0.05	&	0.029	&	26.3	&	0.09	&	26.1	&	0.14	&	26.7	&	0.63	\\
H	&	0.98	&	37.76	&	0.07	&	0.017	&	26.7	&	0.12	&	26.5	&	0.21	&	26.2	&	0.34	\\
I	&	0.28	&	37.26	&	0.19	&	0.005	&	26.8	&	0.18	&	26.4	&	0.24	&	26.0	&	0.44	\\
J	&	0.38	&	37.42	&	0.17	&	0.006	&	---	&	15.64	&	---	&	---	&	---	&	---	\\
K	&	1.33	&	38.00	&	0.10	&	0.023	&	23.3	&	0.01	&	22.9	&	0.01	&	22.7	&	0.01	\\
L	&	0.73	&	37.78	&	0.13	&	0.012	&	26.7	&	0.16	&	26.8	&	0.25	&	---	&	---	\\
M	&	0.40	&	37.55	&	0.23	&	0.007	&	26.7	&	0.15	&	26.9	&	0.29	&	31.5	&	47.57	\\
N	&	0.45	&	37.60	&	0.16	&	0.008	&	23.2	&	0.01	&	22.9	&	0.01	&	22.6	&	0.03	\\
O	&	1.03	&	37.96	&	0.09	&	0.018	&	26.7	&	0.06	&	26.5	&	0.10	&	27.3	&	0.44	\\
P	&	0.63	&	37.75	&	0.11	&	0.011	&	23.7	&	0.03	&	23.3	&	0.02	&	23.1	&	0.03	\\
Q	&	0.73	&	37.81	&	0.09	&	0.013	&	24.8	&	0.04	&	24.6	&	0.04	&	24.3	&	0.07	\\
Total	&	58.13	&	39.54	&	0.07	&	1.000	&												\\
Diffuse	&	28.50	&	39.23	&	0.14	&	0.490	&												\\

\\\\															

\underline{NGC4455}	&&&&\\														
A	&	41.46	&	38.73	&	0.01	&	0.083	&	23.1	&	0.03	&	22.9	&	0.03	&	22.9	&	0.04	\\
B	&	68.44	&	38.95	&	0.01	&	0.137	&	21.6	&	0.01	&	21.3	&	0.01	&	21.2	&	0.01	\\
C	&	11.83	&	38.19	&	0.03	&	0.024	&	22.7	&	0.02	&	22.4	&	0.02	&	22.4	&	0.02	\\
D	&	42.57	&	38.74	&	0.02	&	0.085	&	21.6	&	0.02	&	21.2	&	0.02	&	21.0	&	0.02	\\
E	&	4.54	&	37.77	&	0.08	&	0.009	&	21.0	&	0.01	&	20.6	&	0.01	&	20.4	&	0.01	\\
F	&	6.41	&	37.92	&	0.05	&	0.013	&	21.2	&	0.02	&	20.8	&	0.02	&	20.6	&	0.02	\\
G	&	3.23	&	37.62	&	0.09	&	0.006	&	22.3	&	0.02	&	21.9	&	0.02	&	21.7	&	0.02	\\
H	&	24.08	&	38.49	&	0.02	&	0.048	&	22.2	&	0.03	&	21.8	&	0.02	&	21.6	&	0.02	\\
I	&	14.77	&	38.28	&	0.03	&	0.029	&	21.8	&	0.01	&	21.5	&	0.01	&	21.3	&	0.01	\\
J	&	2.80	&	37.56	&	0.09	&	0.006	&	21.8	&	0.02	&	21.5	&	0.02	&	21.4	&	0.02	\\
K	&	2.34	&	37.48	&	0.12	&	0.005	&	22.4	&	0.04	&	22.1	&	0.03	&	21.9	&	0.03	\\
L	&	6.83	&	37.95	&	0.04	&	0.014	&	22.5	&	0.01	&	22.2	&	0.01	&	22.0	&	0.01	\\
M	&	2.62	&	37.53	&	0.09	&	0.005	&	22.1	&	0.03	&	21.8	&	0.03	&	21.7	&	0.03	\\
N	&	17.06	&	38.34	&	0.03	&	0.034	&	23.1	&	0.02	&	22.8	&	0.02	&	22.6	&	0.02	\\
O	&	5.27	&	37.83	&	0.06	&	0.011	&	22.5	&	0.02	&	22.2	&	0.02	&	21.9	&	0.03	\\
P	&	9.86	&	38.11	&	0.04	&	0.020	&	22.2	&	0.02	&	21.8	&	0.02	&	21.6	&	0.02	\\
Q	&	25.03	&	38.59	&	0.02	&	0.050	&	21.3	&	0.01	&	20.9	&	0.01	&	20.7	&	0.01	\\
R	&	4.81	&	37.95	&	0.08	&	0.010	&	22.2	&	0.01	&	21.8	&	0.01	&	21.6	&	0.01	\\
S	&	11.32	&	38.39	&	0.05	&	0.023	&	23.5	&	0.03	&	23.2	&	0.02	&	23.2	&	0.04	\\
T	&	4.23	&	38.02	&	0.08	&	0.008	&	22.3	&	0.02	&	22.0	&	0.02	&	21.8	&	0.02	\\
U	&	4.39	&	38.10	&	0.08	&	0.009	&	21.6	&	0.02	&	21.2	&	0.02	&	21.0	&	0.02	\\
V	&	3.00	&	37.99	&	0.10	&	0.006	&	22.3	&	0.01	&	22.0	&	0.01	&	21.8	&	0.02	\\
W	&	2.94	&	38.03	&	0.10	&	0.006	&	23.0	&	0.03	&	22.5	&	0.03	&	22.3	&	0.03	\\
X	&	3.61	&	38.16	&	0.07	&	0.007	&	22.5	&	0.03	&	22.4	&	0.03	&	22.4	&	0.04	\\
Y	&	4.62	&	38.32	&	0.10	&	0.009	&	23.1	&	0.01	&	22.8	&	0.01	&	22.6	&	0.02	\\
Total	&	500.78	&	39.81	&	0.02	&	1.000	&												\\
Diffuse	&	172.72	&	39.35	&	0.05	&	0.345	&												\\

\\\\

\end{tabular}
\end{table*}

\begin{table*}
\contcaption{Properties of Individual \HII{} Regions}
\label{tab:continued}
\begin{tabular}{ccccccccccc}
	& $f_\HA{}$ & $\log{L_\HA{}}$ & $\Delta\log{L_\HA{}}$ & $\frac{f_\HA{}}{f_\HA{}(tot)}$ & $\mu(g)$ & $\Delta\mu(g)$   & $\mu(r)$ & $\Delta\mu(r)$   & $\mu(z)$ & $\Delta\mu(z)$	\\
    & $\rm10^{-15}erg\,s^{-1}\,cm^{-2}$ & $\log{\rm erg\,s^{-1}}$ & & & \surf{}& \surf{}& \surf{}& \surf{}& \surf{}& \surf{} \\
 \hline\\	
\underline{UGC6181}	&&&&\\														
A	&	0.53	&	37.55	&	0.06	&	0.002	&	24.8	&	0.03	&	24.4	&	0.03	&	23.8	&	0.04	\\
B	&	6.67	&	38.65	&	0.01	&	0.029	&	24.2	&	0.02	&	23.7	&	0.02	&	23.4	&	0.02	\\
C	&	0.49	&	37.52	&	0.10	&	0.002	&	25.4	&	0.04	&	25.1	&	0.05	&	24.7	&	0.06	\\
D	&	6.09	&	38.61	&	0.01	&	0.026	&	23.8	&	0.02	&	23.4	&	0.02	&	23.2	&	0.02	\\
E	&	1.21	&	37.91	&	0.04	&	0.005	&	24.5	&	0.02	&	24.1	&	0.03	&	23.9	&	0.04	\\
F	&	1.23	&	37.91	&	0.04	&	0.005	&	23.1	&	0.02	&	22.7	&	0.02	&	22.5	&	0.02	\\
G	&	2.09	&	38.15	&	0.02	&	0.009	&	22.8	&	0.03	&	22.4	&	0.02	&	22.1	&	0.02	\\
H	&	0.73	&	37.69	&	0.05	&	0.003	&	22.4	&	0.02	&	22.1	&	0.02	&	21.8	&	0.02	\\
I	&	0.42	&	37.45	&	0.10	&	0.002	&	23.3	&	0.01	&	23.0	&	0.02	&	22.8	&	0.02	\\
J	&	0.08	&	36.72	&	0.34	&	0.000	&	24.6	&	0.03	&	24.4	&	0.04	&	24.2	&	0.04	\\
K	&	0.20	&	37.12	&	0.15	&	0.001	&	22.3	&	0.01	&	21.9	&	0.01	&	21.6	&	0.01	\\
L	&	0.23	&	37.18	&	0.15	&	0.001	&	22.3	&	0.01	&	21.9	&	0.01	&	21.6	&	0.01	\\
M	&	1.12	&	37.88	&	0.05	&	0.005	&	22.7	&	0.02	&	22.3	&	0.01	&	22.0	&	0.01	\\
N	&	0.92	&	37.79	&	0.04	&	0.004	&	23.4	&	0.04	&	22.9	&	0.03	&	22.4	&	0.03	\\
O	&	2.08	&	38.14	&	0.02	&	0.009	&	23.4	&	0.03	&	23.0	&	0.03	&	22.6	&	0.02	\\
P	&	3.51	&	38.37	&	0.02	&	0.015	&	22.8	&	0.01	&	22.4	&	0.01	&	22.1	&	0.01	\\
Q	&	0.60	&	37.61	&	0.07	&	0.003	&	24.7	&	0.03	&	24.2	&	0.03	&	24.0	&	0.04	\\
R	&	0.36	&	37.38	&	0.08	&	0.002	&	22.8	&	0.01	&	22.3	&	0.01	&	22.1	&	0.01	\\
S	&	0.98	&	37.82	&	0.04	&	0.004	&	22.5	&	0.01	&	22.0	&	0.01	&	21.7	&	0.02	\\
T	&	1.88	&	38.10	&	0.03	&	0.008	&	22.8	&	0.02	&	22.3	&	0.02	&	22.0	&	0.02	\\
U	&	1.12	&	37.87	&	0.04	&	0.005	&	23.2	&	0.01	&	22.8	&	0.01	&	22.5	&	0.02	\\
V	&	1.03	&	37.84	&	0.04	&	0.004	&	23.0	&	0.02	&	22.6	&	0.02	&	22.3	&	0.02	\\
W	&	0.44	&	37.47	&	0.09	&	0.002	&	22.9	&	0.01	&	22.5	&	0.01	&	22.2	&	0.01	\\
X	&	0.90	&	37.78	&	0.05	&	0.004	&	23.2	&	0.02	&	22.8	&	0.02	&	22.5	&	0.02	\\
Y	&	4.70	&	38.50	&	0.01	&	0.020	&	23.7	&	0.03	&	23.3	&	0.03	&	23.0	&	0.03	\\
Z	&	2.25	&	38.18	&	0.02	&	0.010	&	23.4	&	0.02	&	23.0	&	0.02	&	22.7	&	0.02	\\
AA	&	1.21	&	37.91	&	0.03	&	0.005	&	25.3	&	0.04	&	24.7	&	0.04	&	24.1	&	0.04	\\
AB	&	0.86	&	37.76	&	0.05	&	0.004	&	23.3	&	0.06	&	23.0	&	0.05	&	22.7	&	0.04	\\
AC	&	0.10	&	36.85	&	0.31	&	0.000	&	23.5	&	0.02	&	23.0	&	0.02	&	22.8	&	0.02	\\
Total	&	233.27	&	40.19	&	0.01	&	1.000	&												\\
Diffuse	&	189.25	&	40.10	&	0.02	&	0.811	&												\\

\\\\															
\underline{UGC6151}	&&&&\\														

A	&	0.90	&	37.80	&	0.05	&	0.017	&	24.9	&	0.02	&	24.6	&	0.03	&	24.5	&	0.06	\\
B	&	0.54	&	37.58	&	0.10	&	0.010	&	25.9	&	0.04	&	25.9	&	0.08	&	25.6	&	0.12	\\
C	&	0.87	&	37.79	&	0.03	&	0.017	&	24.3	&	0.03	&	24.0	&	0.03	&	23.8	&	0.05	\\
D	&	1.42	&	38.00	&	0.02	&	0.027	&	24.2	&	0.02	&	23.8	&	0.02	&	23.6	&	0.04	\\
E	&	0.79	&	37.75	&	0.07	&	0.015	&	25.1	&	0.03	&	24.8	&	0.04	&	24.7	&	0.07	\\
F	&	0.79	&	37.74	&	0.06	&	0.015	&	23.7	&	0.01	&	23.3	&	0.01	&	23.0	&	0.01	\\
G	&	0.24	&	37.23	&	0.14	&	0.004	&	24.6	&	0.02	&	24.2	&	0.02	&	24.0	&	0.03	\\
H	&	2.19	&	38.19	&	0.03	&	0.041	&	24.4	&	0.01	&	24.2	&	0.02	&	23.8	&	0.02	\\
I	&	0.71	&	37.70	&	0.08	&	0.013	&	25.1	&	0.03	&	24.9	&	0.04	&	24.7	&	0.07	\\
J	&	0.96	&	37.83	&	0.06	&	0.018	&	24.2	&	0.01	&	23.8	&	0.02	&	23.6	&	0.03	\\
K	&	0.59	&	37.62	&	0.07	&	0.011	&	23.6	&	0.01	&	23.1	&	0.01	&	22.8	&	0.01	\\
L	&	0.72	&	37.70	&	0.06	&	0.014	&	23.1	&	0.01	&	22.6	&	0.01	&	22.3	&	0.01	\\
M	&	0.47	&	37.52	&	0.07	&	0.009	&	23.3	&	0.01	&	22.8	&	0.01	&	22.5	&	0.01	\\
N	&	8.79	&	38.79	&	0.01	&	0.166	&	24.0	&	0.01	&	23.6	&	0.01	&	23.3	&	0.02	\\
O	&	2.04	&	38.16	&	0.03	&	0.039	&	24.7	&	0.01	&	24.4	&	0.02	&	24.1	&	0.03	\\
P	&	1.04	&	37.87	&	0.06	&	0.020	&	25.0	&	0.02	&	24.7	&	0.03	&	24.6	&	0.05	\\
Q	&	0.87	&	37.79	&	0.03	&	0.016	&	25.0	&	0.03	&	24.9	&	0.04	&	24.9	&	0.09	\\
R	&	0.40	&	37.45	&	0.07	&	0.008	&	25.1	&	0.03	&	25.0	&	0.05	&	24.7	&	0.07	\\
S	&	3.53	&	38.40	&	0.02	&	0.067	&	24.7	&	0.02	&	24.4	&	0.02	&	24.3	&	0.04	\\
T	&	4.08	&	38.46	&	0.01	&	0.077	&	25.0	&	0.02	&	24.7	&	0.03	&	24.6	&	0.06	\\
U	&	3.42	&	38.38	&	0.02	&	0.065	&	24.7	&	0.03	&	24.3	&	0.04	&	24.2	&	0.05	\\
V	&	0.92	&	37.81	&	0.03	&	0.017	&	23.9	&	0.02	&	23.5	&	0.02	&	23.2	&	0.03	\\
W	&	0.54	&	37.58	&	0.08	&	0.010	&	23.4	&	0.02	&	23.0	&	0.07	&	22.6	&	0.11	\\
X	&	0.50	&	37.55	&	0.08	&	0.009	&	23.2	&	0.01	&	22.8	&	0.01	&	22.5	&	0.01	\\
Y	&	1.55	&	38.04	&	0.02	&	0.029	&	25.0	&	0.02	&	24.8	&	0.04	&	24.8	&	0.08	\\
Total	&	52.90	&	39.57	&	0.02	&	1.000	&												\\
Diffuse	&	14.06	&	39.00	&	0.08	&	0.266	&												\\

\\\\

\end{tabular}
\end{table*}

\begin{table*}
\contcaption{Properties of Individual \HII{} Regions}
\label{tab:continued}
\begin{tabular}{ccccccccccc}
	& $f_\HA{}$ & $\log{L_\HA{}}$ & $\Delta\log{L_\HA{}}$ & $\frac{f_\HA{}}{f_\HA{}(tot)}$ & $\mu(g)$ & $\Delta\mu(g)$   & $\mu(r)$ & $\Delta\mu(r)$   & $\mu(z)$ & $\Delta\mu(z)$	\\
    & $\rm10^{-15}erg\,s^{-1}\,cm^{-2}$ & $\log{\rm erg\,s^{-1}}$ & & & \surf{}& \surf{}& \surf{}& \surf{}& \surf{}& \surf{} \\
 \hline\\	
										
\underline{UGC5633}	&&&&\\														
															
A	&	1.12	&	37.87	&	0.08	&	0.008	&	29.4	&	1.38	&	22.9	&	0.02	&	22.6	&	0.02	\\
B	&	0.59	&	37.58	&	0.10	&	0.004	&	28.9	&	0.88	&	---	&	---	&	---	&	---	\\
C	&	0.50	&	37.51	&	0.11	&	0.004	&	25.0	&	0.03	&	24.5	&	0.03	&	24.3	&	0.06	\\
D	&	0.76	&	37.70	&	0.10	&	0.005	&	25.1	&	0.03	&	24.7	&	0.04	&	24.7	&	0.10	\\
E	&	1.38	&	37.96	&	0.07	&	0.010	&	25.3	&	0.04	&	24.8	&	0.05	&	24.5	&	0.08	\\
F	&	1.21	&	37.90	&	0.07	&	0.009	&	24.5	&	0.02	&	24.1	&	0.02	&	24.1	&	0.05	\\
G	&	0.59	&	37.59	&	0.10	&	0.004	&	23.6	&	0.01	&	23.2	&	0.01	&	23.0	&	0.02	\\
H	&	4.11	&	38.43	&	0.02	&	0.029	&	23.4	&	0.02	&	22.9	&	0.02	&	22.6	&	0.03	\\
I	&	3.97	&	38.42	&	0.03	&	0.028	&	23.2	&	0.02	&	22.7	&	0.02	&	22.5	&	0.02	\\
J	&	3.29	&	38.33	&	0.04	&	0.024	&	23.7	&	0.02	&	23.3	&	0.02	&	23.0	&	0.02	\\
K	&	1.26	&	37.92	&	0.08	&	0.009	&	25.2	&	0.03	&	24.6	&	0.04	&	24.4	&	0.09	\\
L	&	0.54	&	37.55	&	0.13	&	0.004	&	25.8	&	0.06	&	25.4	&	0.08	&	25.4	&	0.19	\\
M	&	0.99	&	37.81	&	0.09	&	0.007	&	26.7	&	0.12	&	26.6	&	0.20	&	27.1	&	0.72	\\
N	&	2.26	&	38.17	&	0.04	&	0.016	&	24.3	&	0.01	&	23.9	&	0.02	&	23.8	&	0.04	\\
O	&	2.16	&	38.15	&	0.05	&	0.015	&	24.0	&	0.02	&	23.6	&	0.02	&	23.4	&	0.03	\\
P	&	5.61	&	38.57	&	0.03	&	0.040	&	23.6	&	0.02	&	23.0	&	0.02	&	22.7	&	0.02	\\
Q	&	3.69	&	38.38	&	0.03	&	0.026	&	23.3	&	0.03	&	22.8	&	0.03	&	22.6	&	0.03	\\
R	&	2.00	&	38.12	&	0.06	&	0.014	&	24.9	&	0.03	&	24.6	&	0.04	&	24.4	&	0.10	\\
S	&	11.63	&	38.88	&	0.01	&	0.083	&	24.3	&	0.02	&	23.8	&	0.02	&	23.6	&	0.04	\\
T	&	0.92	&	37.78	&	0.11	&	0.007	&	25.7	&	0.06	&	25.6	&	0.09	&	26.0	&	0.32	\\
U	&	3.15	&	38.31	&	0.04	&	0.022	&	25.8	&	0.05	&	25.5	&	0.07	&	25.8	&	0.21	\\
V	&	1.37	&	37.95	&	0.07	&	0.010	&	24.0	&	0.03	&	23.6	&	0.02	&	23.5	&	0.03	\\
W	&	0.64	&	37.62	&	0.10	&	0.005	&	23.8	&	0.04	&	23.4	&	0.03	&	23.3	&	0.04	\\
X	&	2.21	&	38.16	&	0.07	&	0.016	&	24.1	&	0.02	&	23.8	&	0.02	&	23.6	&	0.03	\\
Y	&	0.51	&	37.52	&	0.15	&	0.004	&	24.6	&	0.04	&	24.2	&	0.03	&	24.2	&	0.06	\\
Z	&	0.64	&	37.62	&	0.14	&	0.005	&	25.9	&	0.05	&	25.6	&	0.08	&	25.6	&	0.22	\\
AA	&	1.75	&	38.10	&	0.09	&	0.013	&	23.4	&	0.02	&	23.3	&	0.01	&	23.0	&	0.02	\\
AB	&	2.39	&	38.27	&	0.07	&	0.017	&	23.7	&	0.01	&	26.2	&	0.11	&	27.3	&	0.86	\\
AC	&	0.92	&	37.89	&	0.12	&	0.007	&	26.1	&	0.06	&	23.5	&	0.01	&	23.2	&	0.02	\\
AD	&	0.50	&	37.65	&	0.18	&	0.004	&	23.8	&	0.02	&	23.3	&	0.01	&	23.1	&	0.02	\\
AE	&	0.72	&	37.84	&	0.18	&	0.005	&	23.7	&	0.02	&	24.6	&	0.05	&	24.0	&	0.08	\\
AF	&	0.45	&	37.67	&	0.20	&	0.003	&	25.1	&	0.03	&	24.7	&	0.06	&	24.6	&	0.13	\\
AG	&	0.78	&	37.94	&	0.11	&	0.006	&	25.0	&	0.04	&	---	&	---	&	---	&	---	\\
Bar	&	6.89	&	38.65	&	0.04	&	0.049	&	23.4	&										\\
Total	&	139.86	&	39.96	&	0.02	&	1.000	&	23.7	&										\\
Diffuse	&	75.24	&	39.69	&	0.04	&	0.538	&	26.1	&										\\

\end{tabular}
\end{table*}

\pagebreak



\bsp	
\label{lastpage}
\end{document}